\documentclass[iop,letterpaper]{emulateapj}
\usepackage{times}
\usepackage{courier}
\usepackage{chngpage}
\usepackage{color}
\usepackage{float}
\usepackage{rotating}
\usepackage[fleqn]{amsmath}
\usepackage{relsize}
\usepackage{amssymb}
\usepackage{multirow}
\usepackage{pdfpages}
\usepackage{setspace}
\usepackage{url}
\usepackage{verbatim}
\usepackage{amsmath}
\usepackage[T1]{fontenc}

\slugcomment{Accepted 2016 September 11 to PASP}

\shorttitle{Small NEAs in PTF}
\shortauthors{Waszczak et al.}

\def\pasp{\ref@jnl{PASP}}               

\begin{document}

\title{Small near-Earth asteroids in the Palomar Transient Factory survey: \\
A real-time streak-detection system}

\author{Adam Waszczak$^1$, Thomas A. Prince$^2$, Russ Laher$^3$, Frank Masci$^4$, Brian Bue$^5$, Umaa Rebbapragada$^5$, \\
Tom Barlow$^2$, Jason Surace$^3$, George Helou$^4$, Shrinivas Kulkarni$^2$\\}

\affil{\\$^1$Division of Geological and Planetary Sciences, California Institute of Technology, Pasadena, CA 91125, USA\\
         $^2$Division of Physics, Mathematics and Astronomy, California Institute of Technology, Pasadena, CA 91125, USA: \href{mailto:prince@caltech.edu}{\color{blue}{prince@caltech.edu}}\\
         $^3$Spitzer Science Center, California Institute of Technology, Pasadena, CA 91125, USA\\
         $^4$Infrared Processing and Analysis Center, California Institute of Technology, Pasadena, CA 91125, USA\\
         $^5$Jet Propulsion Laboratory, California Institute of Technology, Pasadena, CA 91109, USA
         }

\begin{abstract}
Near-Earth asteroids (NEAs) in the 1--100 meter size range are estimated to be $\sim$1,000 times more numerous than the $\sim$15,000 currently-catalogued NEAs, most of which are in the 0.5--10 kilometer size range. Impacts from 10--100 meter size NEAs are not statistically life-threatening but may cause significant regional damage, while 1--10 meter size NEAs with low velocities relative to Earth are compelling targets for space missions. We describe the implementation and initial results of a real-time NEA-discovery system specialized for the detection of small, high angular rate (visually-streaked) NEAs in Palomar Transient Factory (PTF) images. PTF is a 1.2-m aperture, 7.3-deg$^2$ field-of-view optical survey designed primarily for the discovery of extragalactic transients ({\it e.g.}, supernovae) in 60-second exposures reaching $\sim$20.5 visual magnitude. Our real-time NEA discovery pipeline uses a machine-learned classifier to filter a large number of false-positive streak detections, permitting a human scanner to efficiently and remotely identify real asteroid streaks during the night. Upon recognition of a streaked NEA detection (typically within an hour of the discovery exposure), the scanner triggers follow-up with the same telescope and posts the observations to the Minor Planet Center for worldwide confirmation. We describe our ten initial confirmed discoveries, all small NEAs that passed 0.3--15 lunar distances from Earth. Lastly, we derive useful scaling laws for comparing streaked-NEA-detection capabilities of different surveys as a function of their hardware and survey-pattern characteristics. This work most directly informs estimates of the streak-detection capabilities of the Zwicky Transient Facility (ZTF, planned to succeed PTF in 2017), which will apply PTF's current resolution and sensitivity over a 47-deg$^2$ field-of-view.
\end{abstract}


\section{Introduction}

A near-Earth asteroid (NEA) is by definition any asteroid with perihelion $q<1.3$ AU and aphelion $Q>0.983$ AU. From the largest NEA (of diameter $D\approx30$ km) down to $D\approx0.5$ km in size---for which the known population is largely complete---the cumulative size-frequency distribution (Figure~1) goes roughly as $N(D)\propto D^{-2}$, where $N(\text{0.5 km})\approx 10^4$. Harris (\citeyear{har08}, \citeyear{har13}) presents these statistics, and describes how the original `Spaceguard' goal to catalog 90\% of all $D>1$ km NEAs was achieved by the mid-2000s, while the current congressional mandate is to find 90\% of all $D>140$ m NEAs by 2020.

The incrementally-decreasing target size in the NEA census has been mostly motivated by risk mitigation. Over the quarter-century that began with our realization of an asteroid's role in the dinosaurs' extinction ({\it e.g.}, \citealp{alv80}) through to our fulfillment of the 1-km Spaceguard goal, the estimated risk of an individual's death from asteroid impact---initially believed comparable to that of a commercial airplane accident---dropped by an order of magnitude. Surveying to the currently recommended $D>140$ m can decrease this risk by yet another order of magnitude \citep{har08}.  
Hence, discovery of $D<100$ m NEAs will contribute only minimally to any further significant reduction in the risk of death to any individual.
However, events like the Tunguska and Chelyabinsk airbursts \citep{bro13} could have caused caused a significant numbers of deaths if the impact parameters had been different and did cause environmental or other damage.
This  suggests that impacts from 10--100 m objects qualify as `natural disasters' that merit advance warning, 
and possibly prevention via space-based manipulation of hazardous NEAs.

However, the size-frequency distribution informing these estimates is uncertain across orders of magnitude in impactor size, and constrained on the small end ($D\lesssim10$ m) by infrasound detections of bolide fluxes (\citealp{sil09}). 
Besides impact mitigation ({\it e.g.}, \citealp{ahr92}; \citealp{lul05}), other space-based activities benefiting from small NEA discoveries include in-situ compositional studies \citep{mue11} and resource utilization \citep{elv14}. NEAs have also been declared a major component of NASA's manned spaceflight program \citep{oba10}. NEA rendezvous feasibility depends critically on mission duration and fuel requirements, these in turn are functions of the NEA's orbit and relative velocity ($\Delta v$) with respect to Earth (\citealp{sho78}; \citealp{elv11}). Robotic missions that have been proposed may facilitate or complement the manned program.
One example was the proposed Asteroid Retrieval Mission (ARM; \citealp{bro10}), which evolved into the NASA Asteroid Redirect Mission.
At the time of the writing of this paper, NASA has chosen
to retrieve a boulder from a larger asteroid, rather than retrieve an entire small asteroid.
Natural temporary capture of meter-scale NEAs into Earth-centric orbits, if confirmed via the discovery of `mini-moons' (\citealp{gra12}; \citealp{bol14}), would present another appealing class of targets.

Cleary, discovery of 1--100 m sized NEAs is motivated by different (and more diverse) applications than those which have driven the census of larger NEAs. The discovery \emph{method} often likewise differs. Most large NEAs were found via the `tracklet' method of linking several serendipitously-observed positions within a night or across several nights. This is the basis of `MOPS'-like detection software ({\it e.g.,} \citealp{den13}), which in its present state is most efficient at detecting NEAs moving slower than $~\sim$5 deg/day \citep{jed13}. Below this rate, an NEA's individual detections are nearly point-like for typical survey exposure times ({\it e.g.}, 30--60s), and sufficiently localized on the sky given typical intra-night pointing cadences ({\it e.g.}, 15--45 minutes). Hazardous NEAs occupy a range of orbits with moderate eccentricities, and so they spend most of their time far from the Earth and Sun, where their sufficiently slow apparent motions allow them to be easily detected with this technique. 
Searches for NEAs with DECam (Allen et al. 2013 \& 2014) using
conventional MOPS techniques on a large telescope appear to be particularly promising for finding small NEAs.
Recent searches using NEOWISE (Mainzer et al. 2014) have also yielded interesting new detection of small NEAs.

In contrast, the method of \emph{streak detection} enables discovery of much smaller and closer ({\it i.e.}, brighter and faster-moving) NEAs. Whereas slower-moving NEAs can be mistaken for main-belt asteroids, streaked asteroids having an angular rate of larger than 1 deg/day are likely to be NEAs. Unlike the tracklet method, discovery via streak detection is possible on the basis of a \emph{single} exposure via recognition of the streak morphology, meaning repeat visits to the same patch of sky are unnecessary and more area can be searched. 
Lastly, NEAs 
that are detected as streaks are typically closer and therefore are 2 to 3 times brighter than those found by the tracklet method when they are tracked non-sidereally, making them more convenient for follow-up from dedicated (including amateur-class) facilities once an approximate orbit is determined.

Survey-scale application of the streak-detection method for NEA discovery was pioneered by \cite{hel79} using photographic plates on the Palomar 18-inch Schmidt telescope in the 1970s. \cite{rab91} was the first to apply this method with CCD detectors in near real-time with the Spacewatch survey. Combining Spacewatch's streaked NEA detections ({\it e.g.}, \citealp{sco91}) with its tracklet-detected NEAs \citep{jed95} produced a debiased NEA number-size distribution \citep{rab00} spanning four orders of magnitude in size (10 km $>D>$ 1 m).
In 2005, Spacewatch initiated a public Fast Moving Object (FMO) program (\cite{mcmillan+05}) that allowed public access to the survey's imagery on the internet so that members of the public could scan images and report detections of streaks.

\begin{figure}[t]
\centering
\includegraphics[scale=1]{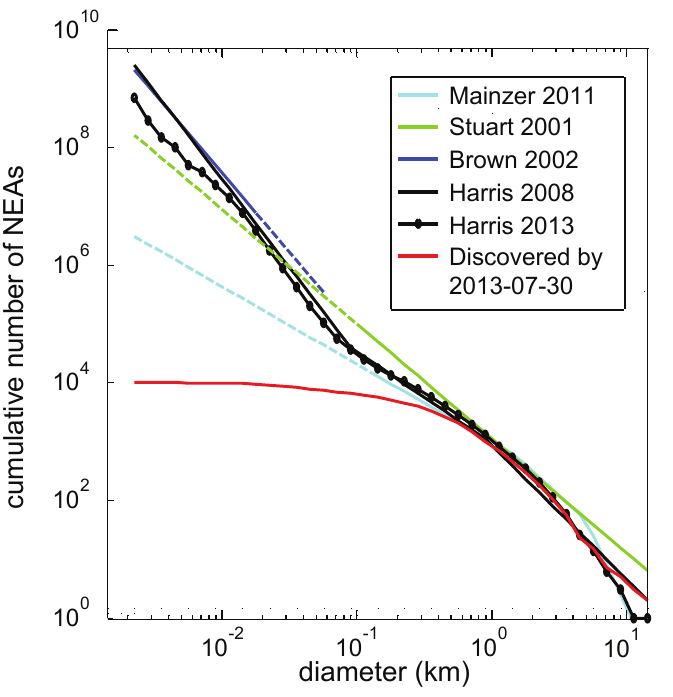}
\caption{Cumulative NEA population distribution models compared to discovered objects. Plot adapted from a figure in \cite{ru14}.}
\end{figure}

\begin{figure}[t]
\centering
\includegraphics[scale=0.85]{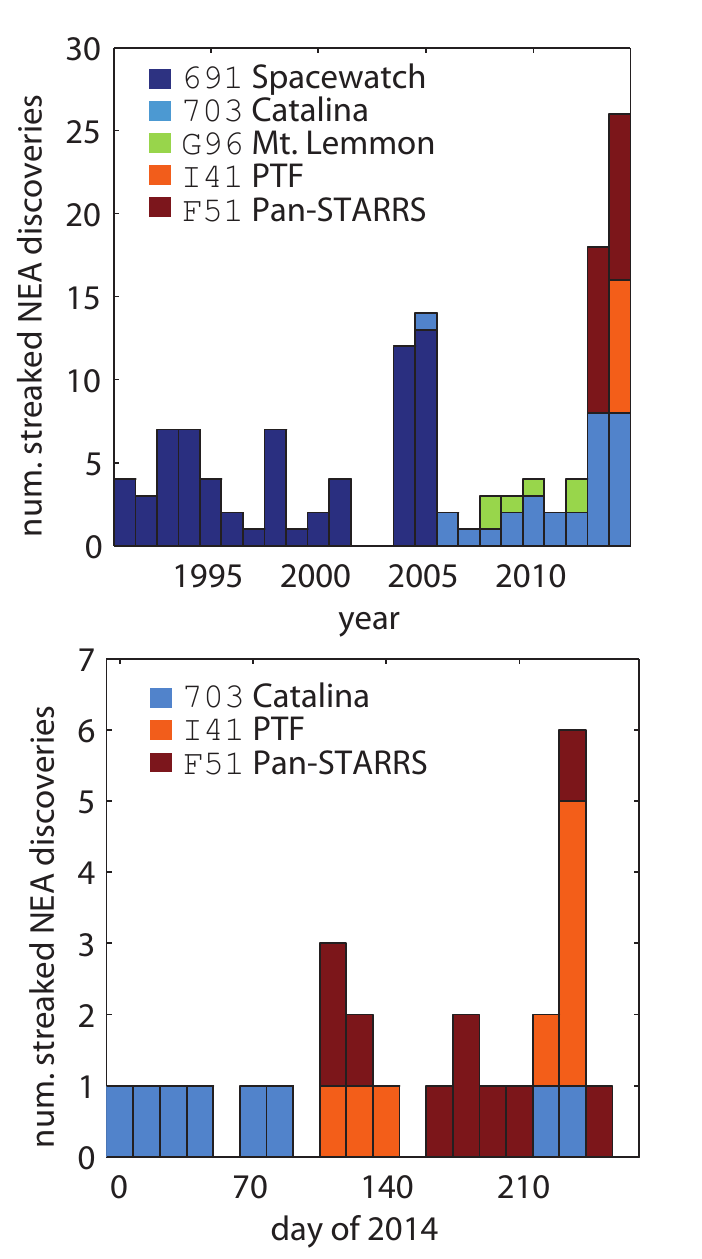}
\caption{Number of streaked NEA discoveries as a function of time (bins include 1990--2014) and survey, where `streaked' is here defined as any discovered streak greater than 10 seeing-widths in length (see text for details).}
\end{figure}

Figure~2 breaks down the number of streaked NEA discoveries as a function of time and survey, from 1991 through 2014-Oct. Here `streaked' is taken to mean any detection wherein the length of the imaged streak is greater than 10 seeing widths. The counts in Figure~2 were compiled by first retrieving all NEA discovery observations from the Minor Planet Center (MPC) database and then using JPL's HORIZONS service \citep{gio96} to compute the on-sky motion at the discovery epoch. These rates were then converted into streak lengths in units of seeing widths, where the continental surveys all have assumed $2''$ seeing and Pan-STARRS has assumed $1''$ seeing. The assumed exposure times come mostly from a table in \cite{lar07}, except for PTF and Pan-STARRS, which have assumed exposure times of 60s and 45s, respectively.

\begin{figure*}[t]
\centering
\includegraphics[scale=0.53]{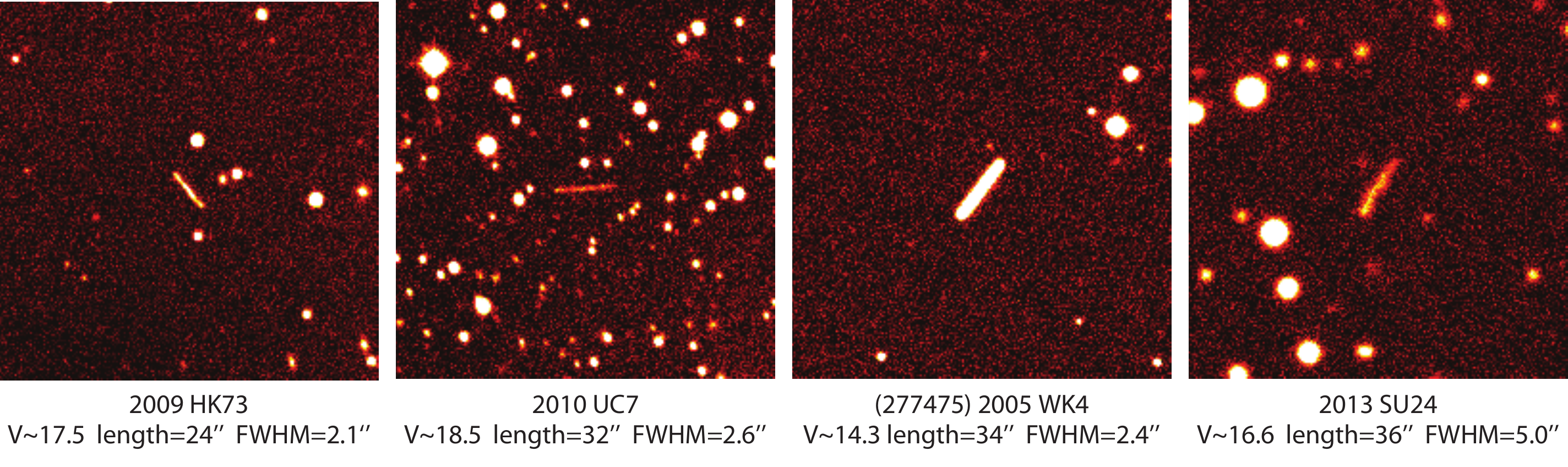}
\caption{Some known small NEAs serendipitously detected by PTF. These observations were retrieved solely by computing these known objects' positions at the epochs of archival PTF images and visually verifying the streak's presence. All images are 200$''$ $\times$ 200$''$ with linear contrast scaling from $-0.5$$\sigma$ to 7$\sigma$.}
\vspace{10pt}
\end{figure*}

Before 2005, Spacewatch was the only contributor of significantly streaked NEA discoveries, and it is also the most prolific streaked-NEA discoverer overall. There are two likely reasons for this: (1) Spacewatch's relatively long 120s exposure time, and (2) the active role of a human screener (`observer') during data collection, as documented by \cite{rab91}. The Catalina Sky Survey also has a dedicated human operator to scan candidates and conduct same-night follow-up \citep{lar07}, which explains its similarly consistent contribution of streaked discoveries. Some major NEA surveys of the past two decades \emph{not} contributing to the streaked discoveries in Figure~2 include LINEAR \citep{sto00}---likely because of its short 8s exposures, as well as NEAT \citep{pra99} and LONEOS \citep{sto02}---which to our knowledge lacked real-time human interaction with their respective data flows. 

The years 2013 and 2014 marked a clear upturn in the discovery of streaked NEAs.
The purpose of this paper is to document a new streak-discovery pipeline which has contributed in part to this increased discovery rate.

\section{Overview of the PTF survey}

\subsection{Technical and operational characteristics}

The \emph{Palomar Transient Factory\footnote{\href{http://ptf.caltech.edu}{\color{blue}{http://ptf.caltech.edu}}}} (PTF) is a synoptic survey designed primarily to discover extragalactic transients (\citealp{law09}; \citealp{rau09}). The PTF camera, mounted on Palomar Observatory's 1.2-m Oschin Schmidt Telescope, uses 11 CCDs (each 2K $\times$ 4K) to image 7.3 deg$^2$ of sky at a time (at $1.0''$/pixel resolution). Most exposures ($\sim$85\%) use a Mould-$R$ filter\footnote{The Mould-$R$ filter is very similar to the SDSS-$r$ filter; see \cite{ofe12} for its transmission curve.} (hereafter ``$R$'') with a 60-second integration time. Science operations began in March 2009, with a nominal 1- to 5-day cadence for supernova discovery and typical twice-per-night imaging of fields. Median seeing is $2''$ with a limiting magnitude $R\approx20.5$ (for 5$\sigma$ point-source detections), while dark conditions routinely yield $R\approx21.0$ \citep{law10}. About 15\% of nights (near full moon) are devoted to an H$\alpha$-band imaging survey of the full Northern Sky.

In January 2013 the PTF project formally entered a second phase called the \emph{intermediate PTF} (`iPTF'; \citealp{kul13}). For most of this paper we simply use `PTF' to mean the entire survey, from 2009 onward, though we note that PTF's NEA-discovery capabilities were conceived, funded, developed and commissioned entirely in this post-2012 `iPTF' period. This is partly because iPTF accommodates more varied `sub-surveys' as opposed to a predominantly extragalactic program, including variable star and solar system science.

As will be detailed later ({\it e.g.}, Figure~15), typical PTF pointings tend to avoid the ecliptic (and hence opposition) in accordance with its primarily non-solar-system science objectives. In recent summer seasons, PTF has also spent the majority of its observing time imaging the dense galactic plane; many such galactic fields contain very high source densities and were not capable of being processed with the streak detection pipeline described below.

\subsection{Previous solar system science with PTF}

The present paper discusses the first NEA-related (and first real-time) work with PTF solar system data; previous PTF solar-system work analyzed archival observations of main-belt asteroids. \cite{pol12} and later \cite{cha14} and \cite{was15} used high-cadence data (which is uncommon in PTF) for `pilot studies' of asteroid rotation lightcurves spanning consecutive nights. \cite{was13} mined PTF for all observations of known asteroids and then searched this data set for activity characteristic of `main-belt comets' \citep{hsi06}. We used this database of known-object observations to extract detections of known streaking NEAs in PTF (Section 2.4). \cite{was13} also developed an original MOPS-like tracklet-finding routine which was later implemented in the real-time IPAC pipeline discussed below, but is otherwise unrelated to the streak-detection pipeline.

\subsection{Real-time data reduction at IPAC}

Since the survey's start, PTF has employed two separate data reduction pipelines serving distinct purposes. A real-time image-subtraction pipeline hosted at the National Energy Research Scientific Computing Center at Berkeley Lab (Nugent et al. in prep.) forms the basis of the extragalactic transient discovery program. A separate, archival-grade image-processing pipeline hosted at the Infrared Processing and Analysis Center (IPAC) at Caltech \citep{lah14} runs during the day and performs flat-fielding, bias-subtraction, source catalog generation, and astrometric and absolute-photometric calibration.

In early 2013, a real-time version of the IPAC image-processing pipeline was put into regular nightly operation. This initial version included daily automated batch submission of main-belt (and slow-moving near-Earth) asteroid observations to the MPC (both known objects and new discoveries). In addition to the above-mentioned image reduction features detailed by \cite{lah14}, the real-time processing includes an original module for image subtraction \citep{mas13}, which uses a deep co-add of $\sim$20 previous PTF images that reaches $V\approx 22$ ({\it i.e.}, a `reference' or `template' image). The reference image is convolved with the new image's PSF kernel prior to subtraction, as described by \cite{mas13}. The creation of this real-time IPAC pipeline precipitated the development of the streak-detection system discussed in this paper.

\subsection{Detections of known streaking NEAs}

Early in the development of our streak detection system, we sought to extract all observations of known fast-moving NEAs from existing PTF data. We used the table of all predicted PTF sightings of all known asteroids compiled by \cite{was13}, updated to include data through early 2014.

There are a total 539 predicted sightings (of 158 unique objects) for which the predicted motion was faster than 10$''$/minute and the predicted magnitude was brighter than $V=20$. For objects having predicted positional uncertainties greater than 10$''$, the images were visually inspected around the predicted location for the presence of a streak. Because the $V<20$ brightness criterion is based upon HORIZONS-predicted magnitudes, which have a typical accuracy of $\sim0.5$ mag, in certain cases the actual magnitude was almost certainly fainter than $V=20$. 
These particular predicted sightings (having good positional localization but possibly too faint for detection) 
are still included as long as the predicted (point-source) magnitude is brighter than $V=20$.

Figure~3 shows some examples of visually-confirmed PTF streak detections from this set of predicted sightings. Qualitative variations in morphology due to a differences in magnitude, streak length and seeing are apparent.

\begin{figure*}[t]
\centering
\includegraphics[scale=0.65]{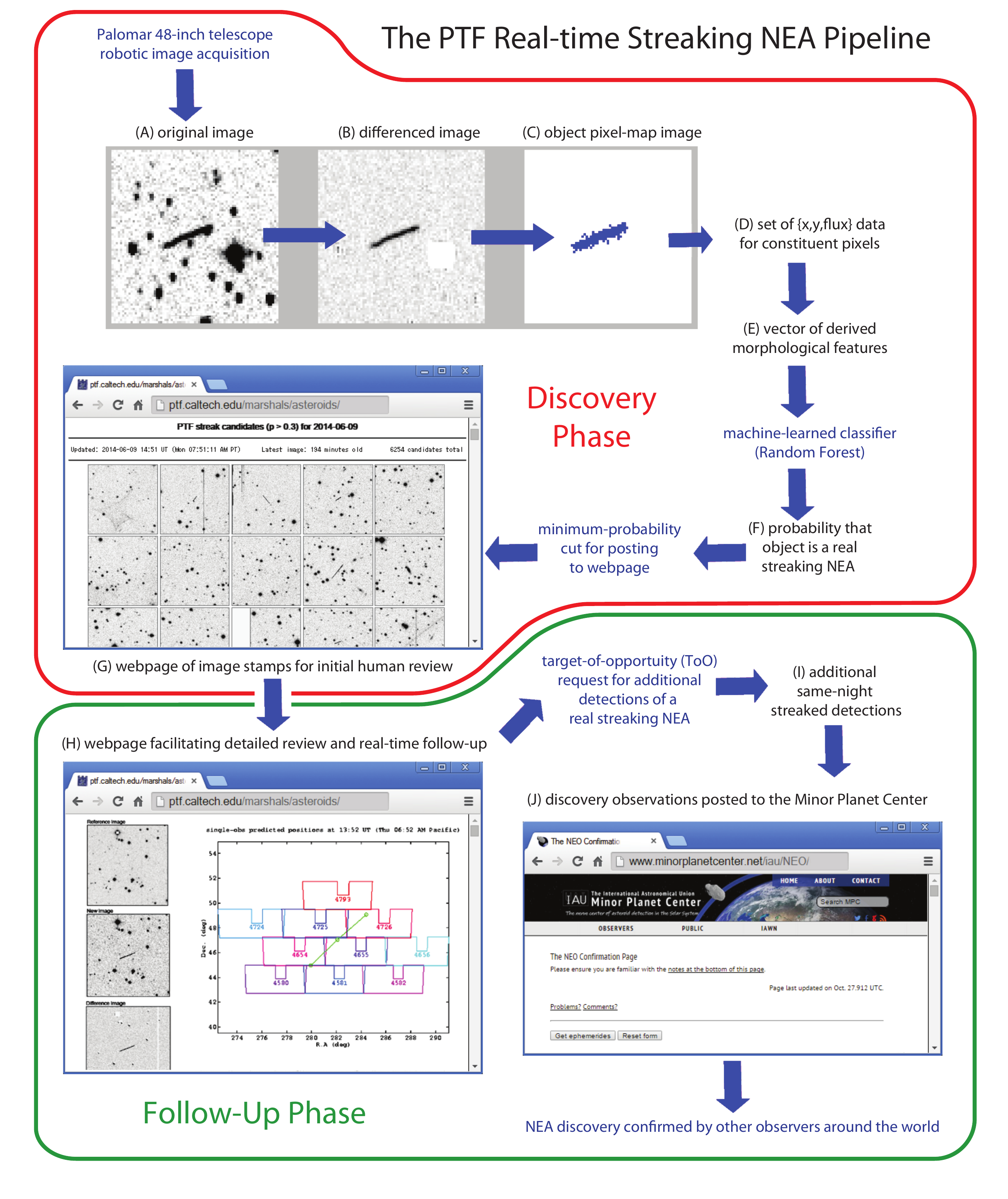}
\caption{Flowchart depicting the PTF streaking-NEA discovery pipeline.}
\end{figure*}

As described below, these 539 serendipitous sightings constituted the initial test bed for development of our streak detection algorithm. 
As of the time of this paper, PTF acquired $\sim$90 new additional detections including unconfirmed PTF discoveries, confirmed PTF discoveries, and PTF-observed discoveries from other surveys in 2014.

\section{Streak-detection process}

The principal steps of the streak-detection process, referenced to component products in Figure~4, are: 

\begin{enumerate}
\item Image processing and subtraction of a reference image to produce a differenced image (A\&B)
\item Detection of candidate streaks as regions of contiguous pixels on the differenced image (C\&D)
\item Measurement of a set of morphological features describing each candidate streak (E)
\item Filtering of likely non-real detections on the basis of their computed features (F)
\item Human recognition of real streaks by reviewing images of the filtered candidates (G)
\end{enumerate}

The above five steps comprise the \emph{discovery} phase and entail the creation of the data products labeled (A) through (G) in Figure~4. Upon discovery of a real streak, data products (H) through (J) are created as part of the \emph{follow-up} phase, which we discuss later (Section 4).

Initial image processing and reference-image subtraction (first of the above-enumerated steps) are described by \cite{lah14} and \cite{mas13}, respectively. Step 2 involves identifying the pixels on the differenced-image belonging to candidate streaks. Whereas pixel-level data for point-source transients ({\it e.g.}, supernovae or slow-moving asteroids) can be efficiently extracted with commonly used software such as \emph{Source Extractor} \citep{ber96}, streaked detections require a distinct approach as their image footprints contain many more pixels, often with much lower signal to noise per pixel. To meet this need we developed an original piece of software called \texttt{findStreaks}.

\subsection{Object detection with \texttt{findStreaks}}

\subsubsection{Algorithm}

The \texttt{findStreaks} software is derived from code originally created for the IPAC processing pipeline to identify and mask very long tracks in PTF exposures due to satellites and aircraft \citep{lah14}. \texttt{findStreaks} was developed in the C programming language to maximize computing speed. The software first thresholds the image pixels above a local background noise level, then groups contiguous pixels into objects or `blobs' ({\it i.e.}, candidate streaks), and lastly computes morphological features for each object.

The findStreaks code is a single-threaded application.  Parts of the
software could be straightforwardly made to process with multiple
threads using the pthread library, but the overall speed-up that 
may result is yet to be quantified. 
The algorithm is likely amenable to GPU parallelism.

Real-time performance is currently achieved on the iPTF system by running as 
many as four pipelines simultaneously on each 24-CPU machine with 
2.4-GHz CPUs and 16 GBytes total memory (parts of the pipeline
other than findStreaks are parallel processes). In this configuration, 
the findStreaks code typically runs in 13-21 seconds per CCD image
for moderately crowded fields, requiring roughly one wall-clock second of run time per
1500 sources extracted from the image by the Source Extractor software.

The differenced-image's local median background values are computed on a coarse grid with 64-pixel grid spacings and 129-pixel windows, with bilinear interpolation used to fill in the pixel values between the grid points. These median values are used to threshold the positive difference image at 1$\sigma$ above the local median. All below-threshold pixels are discarded, and only above-threshold data are considered further ({\it e.g.}, Figure~4 item C). Image-edge pixels are ignored (to avoid artifacts along CCD edges).
Ignoring image-edge pixels does introduce some selection effects, however we do not attempt to estimate absolute rates of NEA detections in this paper.

The \texttt{findStreaks} module arranges all contiguous blobs of pixels, each blob in one or more segments of computer memory, where adjacent pixels in the cardinal and diagonal positions are considered to be connected. For efficient memory management, the module is configured to handle up to 1 million memory segments, and up to 1000 pixels per segment. The sky-background-subtracted blob flux and instrumental magnitude are computed, along with their respective uncertainties. The median and dispersion of the pixel-blob intensity data are computed and subsequent morphological analysis is done only on pixels with intensities that are within $\pm$3$\sigma$ of the median, where $\sigma$ is given by half the difference between the 84th and 16th percentiles. A line is fit to all pixel positions in each blob, and the slope and $y$-intercept are obtained, as well as the linear correlation coefficient:

\begin{equation}
r=\frac{\sum_i (x_i - \bar{x})(y_i - \bar{y})}{\sqrt{(\sum_i (x_i - \bar{x})^2)(\sum_i (y_i - \bar{y})^2)}}.
\end{equation}

\begin{table}
\footnotesize{
\caption{Morphological and other features saved for streak candidates.}
\begin{tabular}{ll}
\hline
feature & description \\
\hline
\texttt{pixels} & number of pixels associated with detected object \\
\texttt{length} & long axis length \\
\texttt{hwidth} & half-width\\
\texttt{dMax} & perp. distance of maximum-flux pixel from longest axis \\
\texttt{angle} & proper angle (in RA, Dec coords) \\
\texttt{median} & median pixel flux\\
\texttt{scale} & 1$\sigma$ variation in pixel flux\\
\texttt{slope} & slope ($dy/dx$ in image coordinates) of fitted line \\
\texttt{correl} & correlation coefficient of fitted line\\
\texttt{flux} & total flux of object\\
\texttt{refDist} & distance from midpoint to nearest object in reference image\\
\texttt{refMag} & magnitude of nearest object in reference image\\
\texttt{epoch} & epoch (modified Julian date)\\
\texttt{ra} & right ascension of object midpoint\\
\texttt{dec} & declination of object midpoint\\
\hline
\end{tabular}
\bigskip
}
\end{table}

\begin{figure*}[t]
\centering
\includegraphics[scale=0.8]{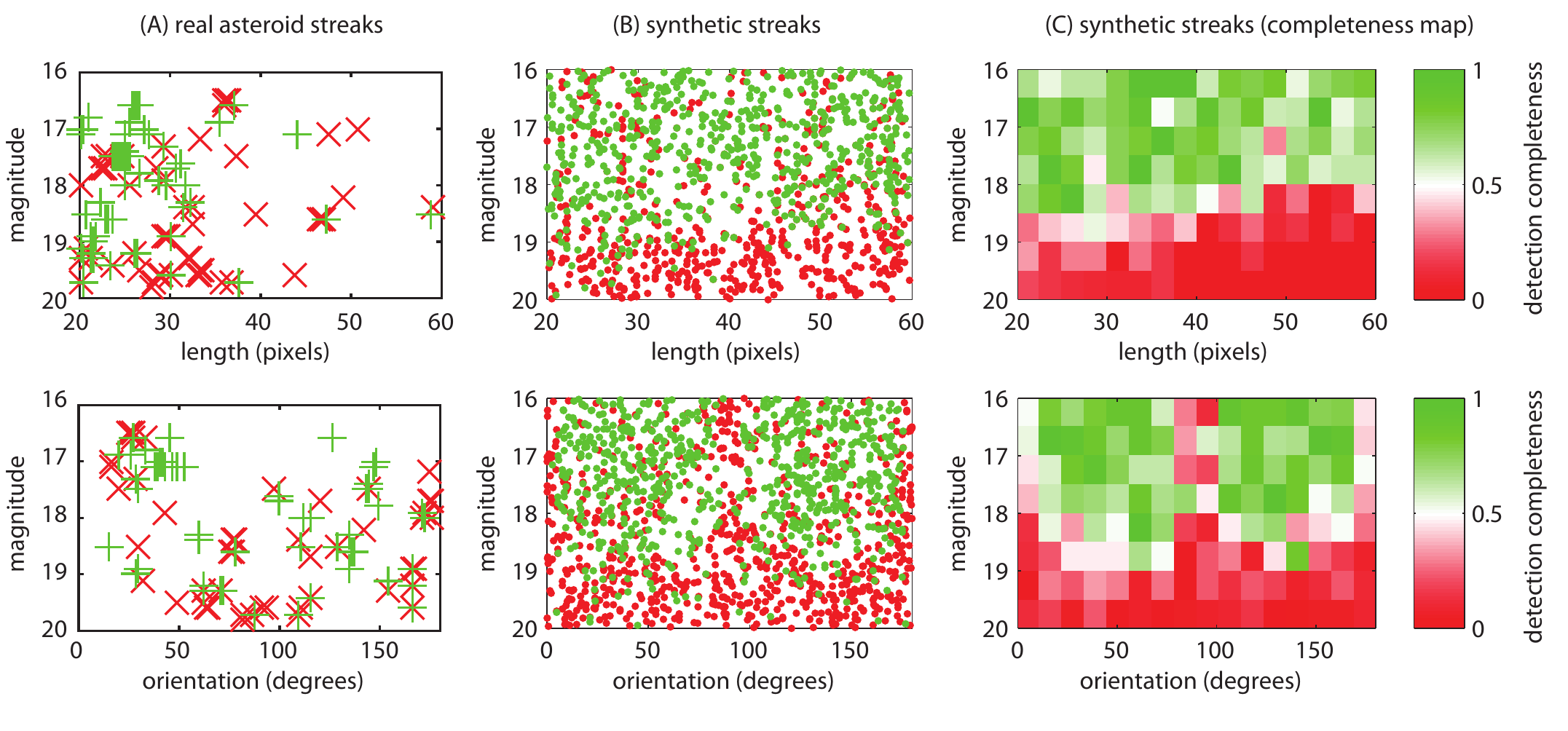}
\caption{Successful detections (green) and failed detections (red) for both real asteroids and synthetic streaks. Here a `successful detection' means an object was found by \texttt{findStreaks} at the predicted location having a measured length within four streak-widths of the predicted length. In the real data (leftmost plots), multiple detections of unique objects are often very close to one another in the 2D spaces plotted here, such that the total number of points discernible on the plot may appear less than actual.}
\end{figure*}

\begin{figure}
\centering
\includegraphics[scale=0.8]{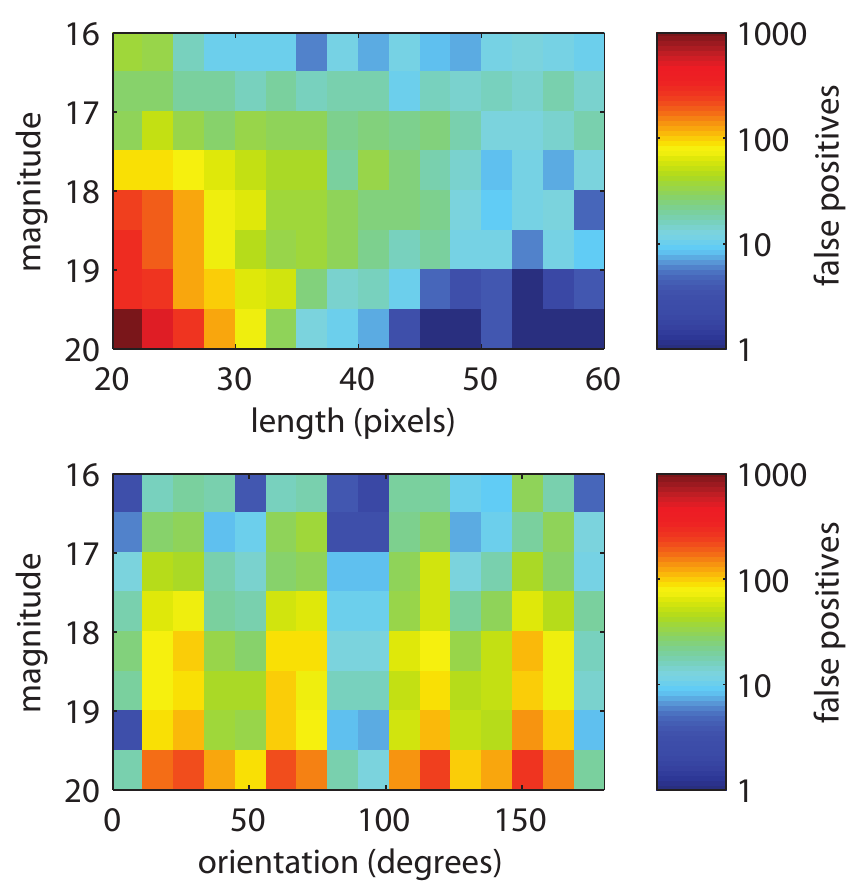}
\caption{Distribution of false positive detections from \texttt{findStreaks}. The largest concentration of these `bogus' detections are in the short and faint regime. Structure as a function of orientation angle (bottom) is due to a combination of the correlation sensitivity (see text) and pixel effects, wherein diagonal ($\pm$$45^{\circ}$-oriented) blobs are less likely to exist as their flux is diluted across more pixels.}
\end{figure}

Perpendicular distances from the linear model to constituent pixels are used to find the blob half-width, defined as half the difference between the 84th and 16th percentiles of these distances. With size and shape parameters now in hand, several hard filters are used to eliminate blobs that are not considered to be streaks. Blobs containing more than 400,000 pixels, having long axes shorter than 9 pixels ({\it i.e.}, 3.6 deg/day motion), or having half-widths larger than 16 pixels are discarded. Blobs for which the absolute value of the linear correlation coefficient is less than 0.5 are also discarded. 

The \texttt{findStreaks} module outputs a table of streak metadata, where each table row corresponds to a streak detection. The real-time pipeline augments this table with additional columns including the proximity of the candidate to the nearest reference-image (stationary) object, as well as the brightness of this nearest reference-image object. Table~1 lists the 15 features currently retained for each candidate, and used in the classification stage that follows. This list of features will be updated to include additional morphological metrics in future versions of this software, but the results of this paper only include analysis of the above-described 15 features.

\subsubsection{Completeness and contamination}

To ascertain \texttt{findStreaks}'s completeness and the number and nature of false positives it detects, we tested the software on a set of images containing both known real asteroid streaks (the 539 predicted sightings described in Section 2.4) and a large number of injected synthetic (simulated) streaks.

To generate each synthetic streak `stamp', we first considered a 2D-Gaussian point-spread function of flux $f$, full-width at half-maximum (FWHM) $\theta$, and center at $(x_0,y_0)$:

\begin{equation}
\begin{aligned}
&\text{PSF}(x,y,x_0,y_0,f,\theta)=\\
&f\times \dfrac{4 \ln 2}{\pi \theta^2} \times \exp\left(-\dfrac{(x-x_0)^2+(y-y_0)^2}{\theta^2} \times 4\ln 2\right)
\end{aligned}
\end{equation}

\noindent In terms of Eq. (1), a simulated asteroid streak of length $L$ oriented at angle $\phi$ is given by

\begin{equation}
\begin{aligned}
\text{Streak}&(x,y,x_0,y_0,f,\theta,L,\phi)=\\
&\frac{1}{L}\int_{t=0}^{t=L}\text{PSF}(x-t \cos\phi,y-t\sin\phi,x_0,y_0,f,\theta)\;dt
\end{aligned}
\end{equation}

\begin{figure*}[t]
\centering
\includegraphics[scale=0.585]{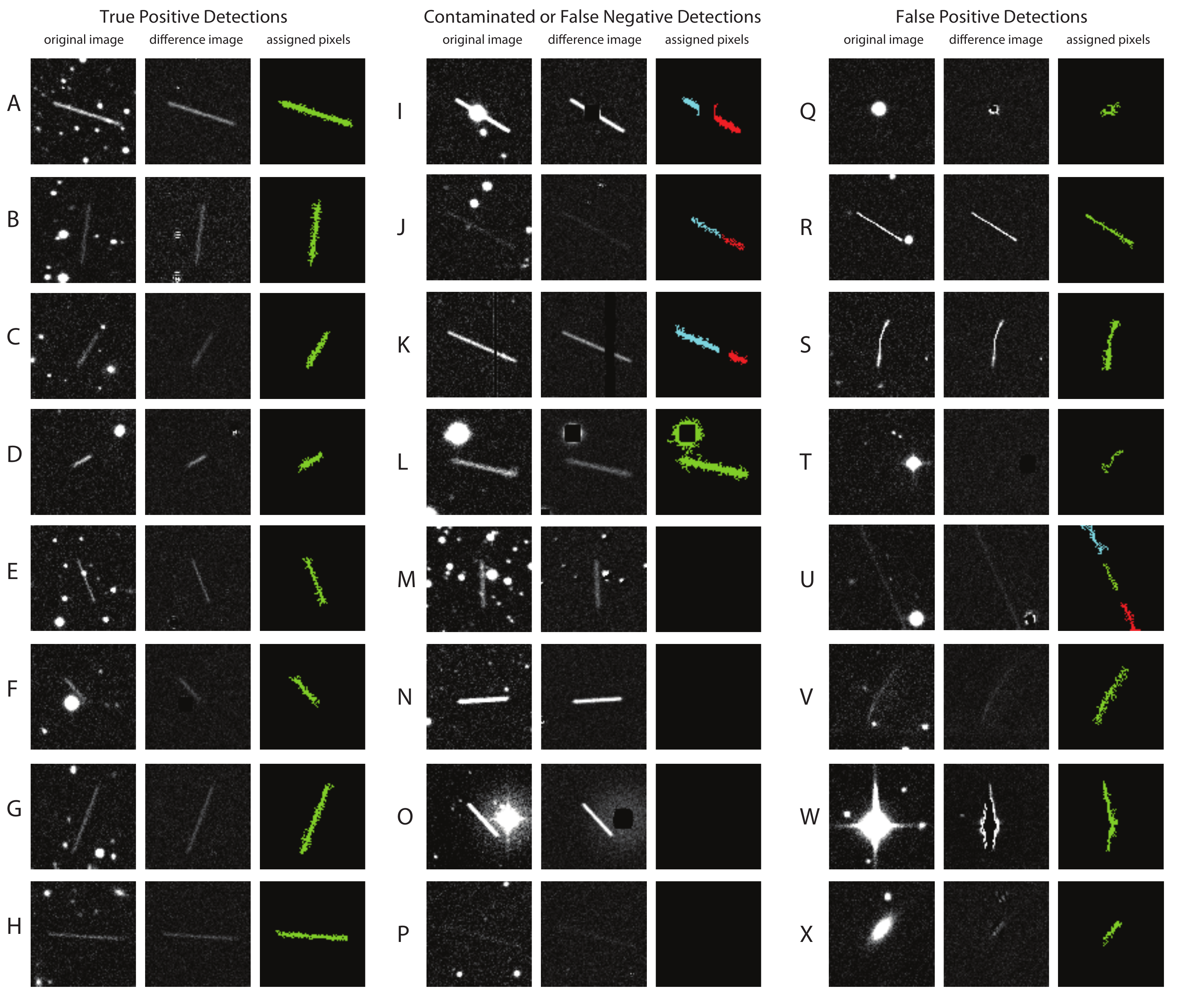}
\caption{Examples of streak detections in PTF images. The third column, "assigned pixels", shows the pixels mapped to the object by \texttt{findStreaks}, wherein unique objects are distinctly colored. (I) Splitting due to saturated star (undefined pixels on difference image). (J) Splitting due to faintness. (K) Splitting due to bad column in difference image. (L) Extraneous pixels from nearby bright star halo. (M) Missed detection due to near-vertical orientation. (N) Missed detection due to near-horizontal orientation. (O) Missed detection due to large variation in background levels (star halo). (P) Missed detection due to faintness. (Q) Poorly-subtracted star false-positive. (R) Linear radiation hit. (S) Non-linear radiation hit. (T) False positive due to background noise. (U) Isolated segment of longer faint streak ({\it e.g.}, due to a satellite). (V) Portion of optical ghost artifact. (W) Diffraction spike false positive. (X) Poorly-subtracted galaxy false-positive.}
\end{figure*}

\noindent \cite{ver12} presents a similar streak model albeit with a slightly different analytical expression.

We evaluate the integral in Eq. (2) numerically over a grid with spacings $\Delta x=\Delta y=0.05''$. Assuming the physical units of $x$ and $y$ are PTF-image pixels (= 1.0$''$), and assuming a typical PTF seeing value of $\theta\approx 2''$ (though we randomly vary $\theta$ along with other parameters, see below), the 0.05$''$ grid spacing ensures the simulated streak is initially oversampled (by a factor of several tens) relative to the final (coarsened) image of the streak.

For each synthetic streak, the various model parameters in Eq. (2) are randomly drawn from flat distributions on the following intervals:

\begin{equation}
\begin{aligned}
0&<x_0<1 \;\;\;\;\;\;\;\;\;\;\;\;\;\; 0<y_0<1\\
1.4''&<\theta<3'' \;\;\;\;\;\;\;\;\;\;\; 10''<L<60''\\
0^{\circ}&<\phi<180^{\circ} \;\;\;\; \text{1800 counts} <f< \text{7200 counts}.
\end{aligned}
\end{equation}

\noindent A synthetic streak's flux $f$ relates to its apparent magnitude $m$ according to $m=m_0-2.5\log_{10}f$, where $m_0$ is the zeropoint of the streak's host image. As the insertion of synthetic streaks into host PTF images is random (see below), it follows that the apparent magnitude $m$ is not sampled from a \emph{uniform} distribution, unlike the parameters $f$, $\theta$, $\phi$ and $L$. The counts for $f$ prescribed above roughly simulate 15 mag $< V<$ 21 mag for typical PTF zeropoints (given normal variations in sky background, extinction, etc.).

To coarsen each synthetic streak (prior to injection into an image), we evaluate the mean flux value in each $1''\times 1''$ bin, equivalent to downsampling the initial simulated image by a factor 20. We round the counts in each resulting pixel to the nearest integer, and crop the streak image to a rectangular `stamp' including all non-zero pixels. Lastly, to simulate shot noise, we replace the value of each non-zero pixel with a random integer sampled from a Poisson distribution whose mean is equal to the original pixel value.

We generated a set of 5,000 synthetic-streak image stamps following the above process, and then inserted these at random locations into the 539 PTF images containing each of the 539 predicted known streak sightings (Section 2.4). In particular, for each image the number of streaks injected was determined by drawing from a Poisson distribution with mean equal to 5. Our results are not highly sensitive to the number of injections.
A set of that number of synthetic streaks was then randomly drawn (with replacement) from the pool of 5,000 and stamped into the image at randomly-chosen $(x,y)$ coordinates. The total number of injected streaks was 2,631. We then processed each image with an offline version of the IPAC real-time image-differencing pipeline (Section 2.3) and ran \texttt{findStreaks} on the differenced images.

In the real-time streak detection pipeline, the output of \texttt{findStreaks} is subsequently subjected to machine-learned classification and human vetting. However, in the interest of initially assessing the completeness and reliability of \texttt{findStreaks} as an isolated module, we here simply (albeit arbitrarily) define a `successful detection' ({\it i.e.} a true positive detection) as any case wherein \texttt{findStreaks} found an object whose measured center lies within a 15$''$ radius of the streak's true center, and the measured length minus the true length is less than four times the streak's measured width. The successful and failed detections according to these criteria are plotted in Figure~5. 

Two trends evident from the synthetic streaks are limiting magnitude-vs.-length (Figure~5 top row) and lack of sensitivity to near-vertical or near-horizontal streaks (Figure~5 bottom row). In general the completeness drops sharply at a certain limiting magnitude; this limiting magnitude brightens from $\sim$19 mag at 20 pixels to $\sim$18 mag at 60 pixels. Streaks oriented very near to either 0$^\circ$ = 180$^\circ$ or 90$^\circ$ are much less reliably detected by \texttt{findStreaks} (at all magnitudes)---this is due the imposed hard limit on correlation ($|r|>0.5$), a criterion which both near-vertical and near-horizontal streaks fail to satisfy.
These are due, for instance, to diffraction spikes and CCD imperfections. See additional discussion below in section \ref{sec:ml}.

The total number of candidate streaks returned by \texttt{findStreaks} in this test was 21,783, or an average of $\sim$40 per image, most being false positive detections (also referred to as `bogus' detections later in this paper). Figure~6 details the distribution of false positives in magnitude, length and orientation space. These plots indicate that the most common type of false-positive detections are faint and short, consistent with these contaminants being mostly star/galaxy subtraction artifacts and segments of extended, low-surface brightness objects like optical ghosts, space debris trails and bright-star halos. Figure~7 presents a gallery of examples of successful, failed, and contaminant detections.

Among the 539 predicted sightings of real streaking asteroids (see section 2.4) in the test images, a total of 240 were successfully detected by \texttt{findStreaks}. The left-side plots of Figure~5 show successful and failed detections in the same feature subspaces in which the synthetics are also plotted in Figure~5. A distinction between the $y$-axis-plotted `magnitude' for the reals and that of the synthetics is that the magnitude of the reals is again the \emph{predicted} brightness, accurate to $\sim$0.5 mag, whereas the synthetic magnitudes are more precisely known (even for the non-detected synthetics, as they still have a known flux and well-defined image zeropoint).

While we have characterized the completeness and contamination of the \texttt{findStreaks} algorithm, we have not undertaken a detailed optimization of the algorithm to maximize the number of real streaking 
asteroids detected by the algorithm. 
Future upgrades of \texttt{findStreaks} are planned, particularly for implementation of a small asteroid detection pipeline for the Zwicky Transient Facility (see Section \ref{sec:scaling} for a discussion of ZTF ).
\newpage
\subsection{Machine-learned classification}
\label{sec:ml}
\begin{figure}
\centering
\includegraphics[scale=0.38]{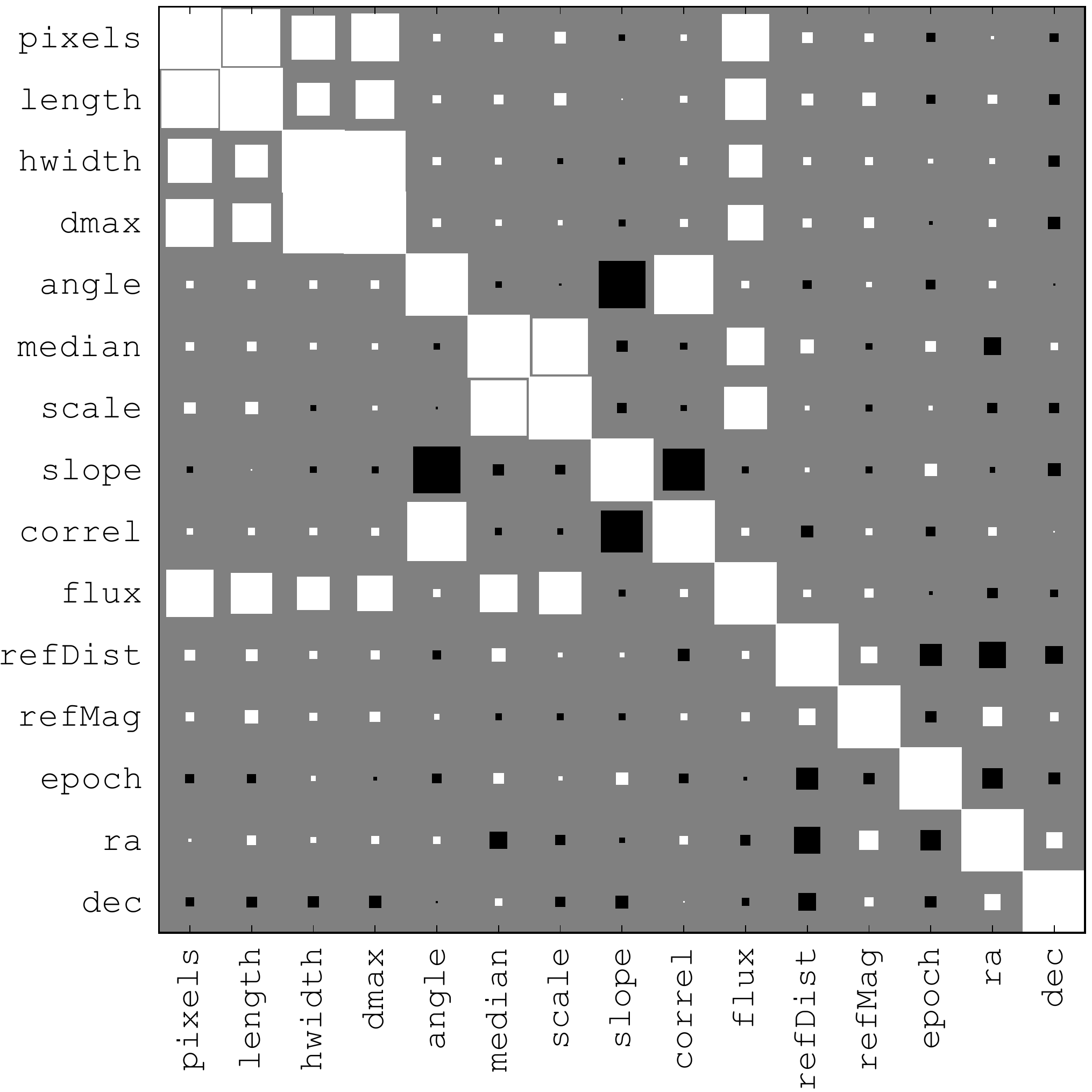}
\caption{Correlation matrix for the 15 features (descriptions given in Table~1) used in the classification process. White squares indicate positive correlation, black indicate negative (anti-) correlation, and the area of each square indicates the magnitude of the correlation.}
\end{figure}

\begin{figure}
\centering
\includegraphics[scale=0.21]{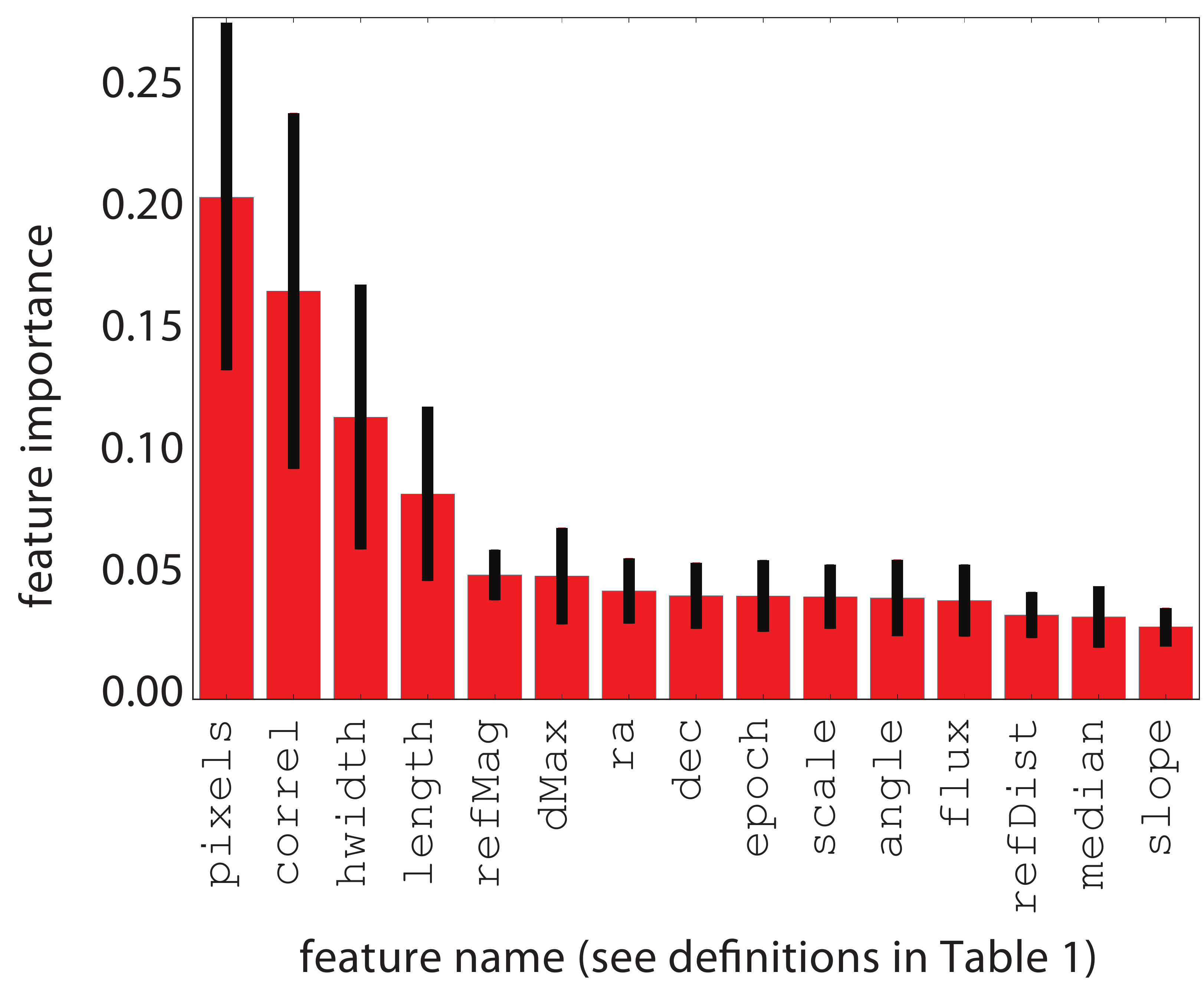}
\caption{Importance of each of the 15 features (descriptions given in  Table~1) used in the classification process. This number represents the fraction of training samples in which each feature contributes more by virtue of being at an earlier node splitting in the decision tree.}
\end{figure}

\subsubsection{Overview}

As described above, a typical PTF image (single CCD) may contain several tens of false positive streak candidates, so that a full night of PTF observations---consisting of several thousand such images---may typically produce of order $10^5$ raw candidate objects. This is far too many to screen by eye. 
This is especially important for small NEAs with high angular rates that can move quickly enough that they may not be in the observed field during the second observation
of a night.
As discussed in Section \ref{sec:too}, if a streaked NEA candidate is identified in a single image, ``target of opportunity'' observations are made later in the night of those fields that 
may contain the NEA based on streak length and orientation.

In addition to reducing the number of candidates for human vetting, there is a second advantage to machine-learned classification.
Imposing simple filters on the measured morphological features (Table~1) can eliminate large subsets of false positives, but these hard cuts generally come at the cost of decreased completeness. 
A good example is the filter on the linear correlation coefficient condition ($|r|>0.5$) discussed above, and the resulting insensitivity to near-vertical and near-horizontal streaks.

\begin{figure*}
\centering
\includegraphics[scale=0.43]{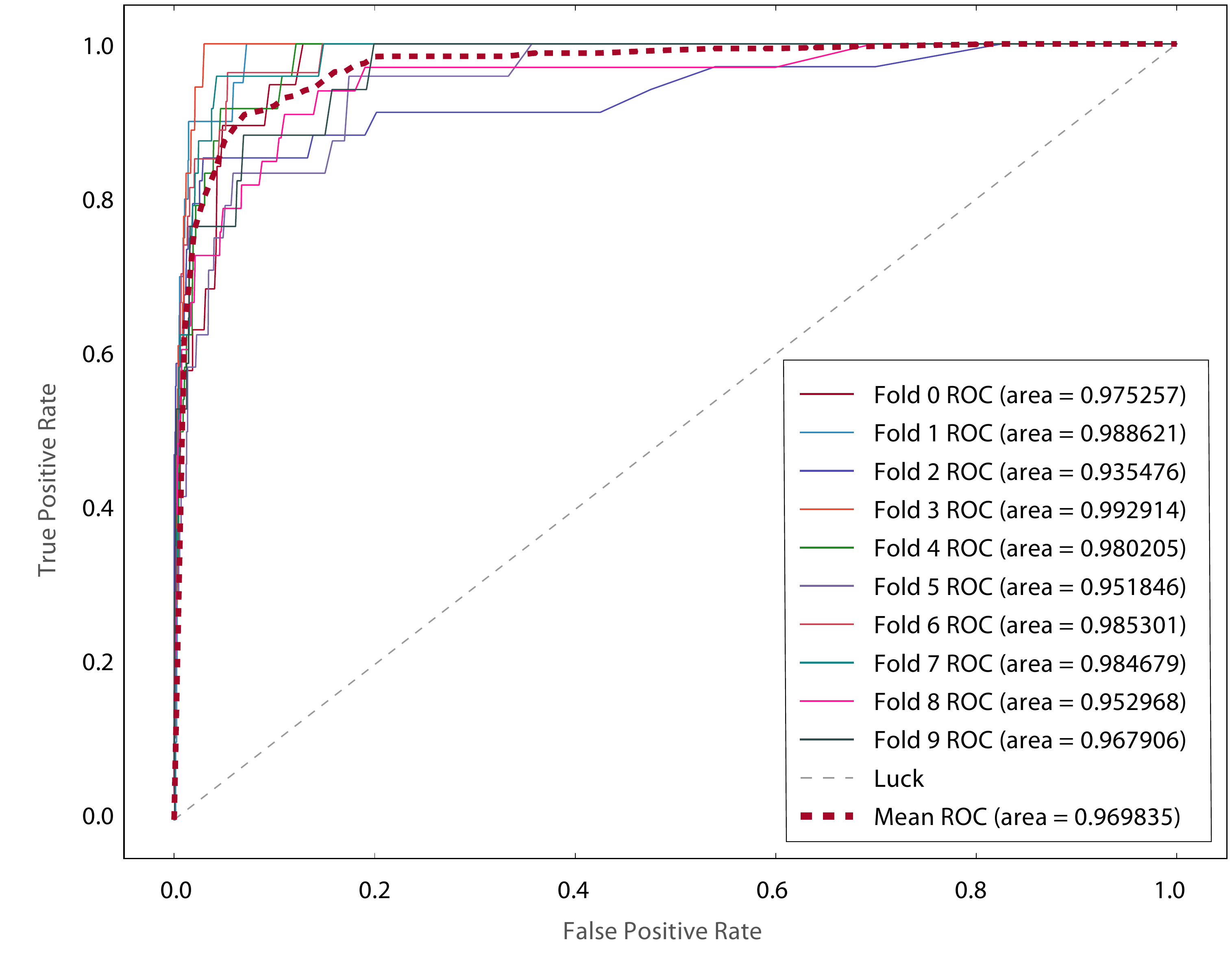}
\caption{Classifier performance for each of the ten cross-validation trials. A plot in true-positive versus false-positive space is commonly referred to as a \emph{receiver operating characteristic} (\emph{ROC}) curve.}
\end{figure*}

To address this issue we have trained and implemented a machine-learned classifier to discriminate real streaks from false positives. We adopt a supervised ensemble-method approach for classification, originally popularized by \cite{bre84}, specifically the \emph{random forest} (RF) method \citep{bre01}. RF classification has extensive and diverse applications in many fields ({\it e.g.}, economics, bioinformatics, sociology). Within astronomy in particular RF classification is one of the more widely-employed methods of machine-learning, though many alternatives exist. For example, \cite{mas14} use the RF method for variable-star lightcurve classification, while others have approached this problem via the use of, {\it e.g.}, support vector machines \citep{woz04}, Kohonen self-organizing maps \citep{bre04,masters+15}, Bayesian networks and mixture-models \citep{mah08}, principlal component analysis \citep{deb09}, multivariate Bayesian and Gaussian mixture models \citep{blo11}, and thick-pen transform methods \citep{par13}.

For general descriptions of RF training and classification, we refer the reader to \cite{bre01}, \cite{brem04}, and the many references cited by \cite{mas14}. Our use of a RF classifier is particularly motivated by its already-proven application to the discovery and classification of astrophysical transients in the same PTF survey data \citep{blo12}.

Streak candidates in PTF images are cast into a vector of quantitative morphological and contextual \emph{features}, namely the 15 features listed in Table~1. Given a large set of such candidates, these metrics define a multi-dimensional space, which can be hierarchically divided into subspaces called \emph{nodes}. The smallest node---also known as a \emph{leaf}---is simply an individual candidate. Given a set of leaves with class labels, {\it i.e.}, a \emph{training set}---one can build an ensemble of decision trees (called a \emph{forest}), each tree representing a different, randomly-generated partitioning of the feature space with respect to a subset of the total training sample (and a subset of the total list of features). The forest allows one to assign a probability that a given vector of features belongs to a given class. For the PTF candidates, we are interested in a binary classification, {\it i.e.}, whether the candidate is real or `bogus'. \cite{blo12} and \cite{brink+13} coined the term \texttt{realBogus} to describe this binary classification probability. In the present work we are essentially adapting Bloom et al.'s \texttt{realBogus} concept to the problem of streaking asteroid discovery.

\subsubsection{Implementation and training}

We employ a Python-based Random Forest classifier included as a part of the \texttt{scikit-learn} Python package\footnote{\href{http://scikit-learn.org/}{\color{blue}{http://scikit-learn.org/}}}. Specifically, we use the \texttt{ExtraTreesClassifier} class in the \texttt{sklearn.ensemble} module. This particular code is an implementation of the `extremely randomized trees' method \citep{geu06}, a variant of the Random Forest method containing an added layer of randomness in the way node-splitting is performed. Specifically, ExtraTrees chooses thresholds randomly for each feature and picks the best of those as the splitting rule, as opposed to the standard RF which picks thresholds that appear most discriminative. The additional randomization tends to improve generalization over the standard RF algorithm, this was verified empirically for our streak data.

\begin{figure*}[t]
\centering
\includegraphics[scale=0.8]{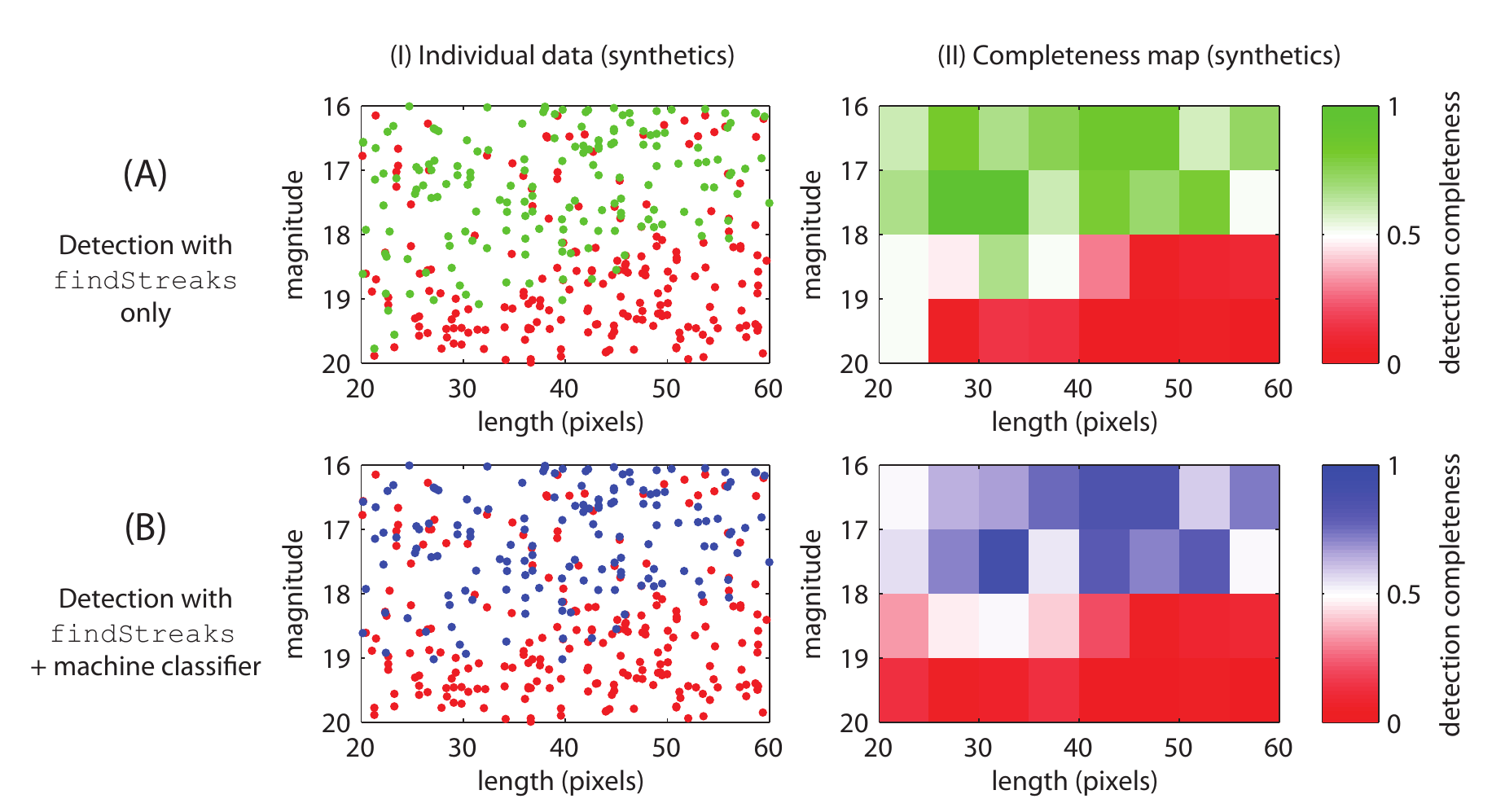}
\caption{In the top row (\texttt{findStreaks} only), detection is again defined as the presence of an object whose length is within four streak widths of the true length, as in Section 3.2.1. In the bottom row (\texttt{findStreaks} plus the classifier), detection is defined as the presence of an object of length within four streak widths of the true length \emph{and} a classification score of $p>0.4$.}
\end{figure*}

\begin{figure}
\centering
\includegraphics[scale=0.8]{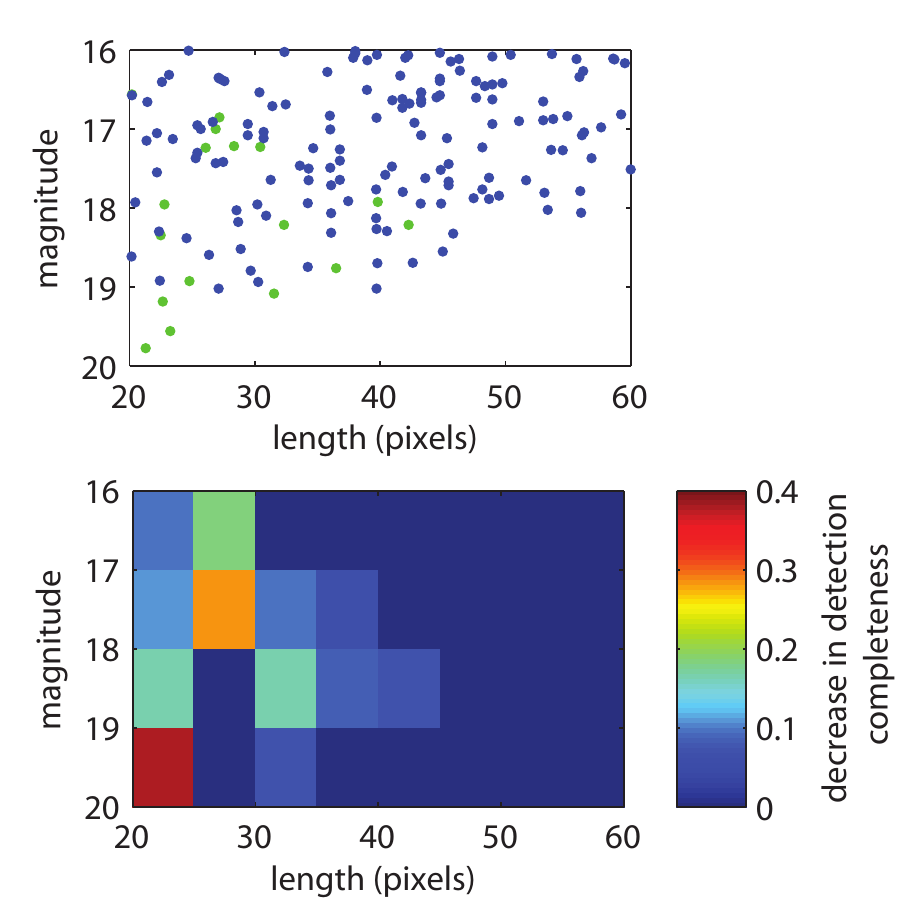}
\caption{Loss in detection completeness due to the machine classifier---{\it i.e.}, like Figure~10 except considering only those candidates that were first positively detected by \texttt{findStreaks}).}
\end{figure}

Our training data consists of all candidate streak detections from the 539-image synthetic-injection test described in Section 3.1.2. This includes 240 real detections (out of the 539 predicted sightings from Section 2.4), 1,285 synthetic detections (out of the 2,631 total injected) and 20,072 bogus detections. Various examples of these bogus (false-positive) detections are shown in the right column of Figure~7, while their distributions in magnitude-vs.-length and magnitude-vs.-orientation space are shown in Figure~6.

Among the 15 features (Table~1) describing the streak candidates, several of the features exhibit some level of correlation, as shown in Figure~8. Most correlations are reasonable as they express the relationship between geometrically-similar quantities: the length of a streak is generally correlated with the number of pixels, and the fitted linear slope correlates with the proper angle. A strong correlation between \texttt{median} and \texttt{scale} (measures of flux signal and noise, respectively) is simply a expression of the Poisson noise associated with photon counting. Assessing the correlation between features aids in the interpretation of relative feature importances (Figure~9) derived during the training process (described below). In particular, among the top four most discriminative features (according to Figure~9), three are significantly correlated (\texttt{pixels}, \texttt{hwidth} and \texttt{length}).

The classifier evaluation consists of a 10-fold bootstrapping process, wherein we split the data (reals, synthetics and boguses) into 10 disjoint sets using stratified random sampling. Then, in each cross-validation fold, we train using 9 of the sets and test on the remaining one---however, we exclude the synthetics from this test sample. In each of the ten cross-validation trials, the classifier outputs a classification probability for each object in the test sample, and we track the true positive rate (TPR; fraction of real streaks accurately classified as reals) as a function of the false-positive rate (FPR; fraction of bogus streaks inaccurately classified as reals). In astrostatisics TPR is also commonly called \emph{completeness} while FPR is equivalently one minus the \emph{reliability}. The results of the separate trials, as well as the averaged result, are shown in Figure~10. By tuning the minimum classification probability ({\it i.e.}, the \texttt{realBogus} score) used to threshold the classifier's output, one effectively moves along the hyperbola-shaped locus of points in TPR-vs.-FPR space seen in the plot.

Several parameters can be adjusted or tuned when working with a random forest classifier. First is the number of decision trees generated during the learning stage. Classification accuracy typically increases with the number of trees and eventually plateaus. Most applications employ hundreds to thousands of trees; here we found that 300 trees provide sufficient performance. 
Increasing the size of the forest to beyond 300 trees did not produce substantially more accurate results.
Another tunable parameter is the number of randomly-selected features (out of the 15 total here considered) with respect to which nodes are split in building the decision trees. \cite{bre01} recommends using the square root of the number of features; however, here we found optimal accuracy when splitting with respect to \emph{all} 15 features. Other parameters that can be tweaked are the maximum depth of a tree, the minimum number of samples per leaf, the minimum number of samples used in a split, and the maximum number of leaf nodes. We do not constrain any of these parameters, meaning we allow: trees of any depth, with any number of leaf nodes, leaf nodes consisting of a single sample, and splits based on the minimum of 2 samples. 

\subsubsection{Post-training performance}

\begin{figure*}[t]
\centering
\includegraphics[scale=0.5]{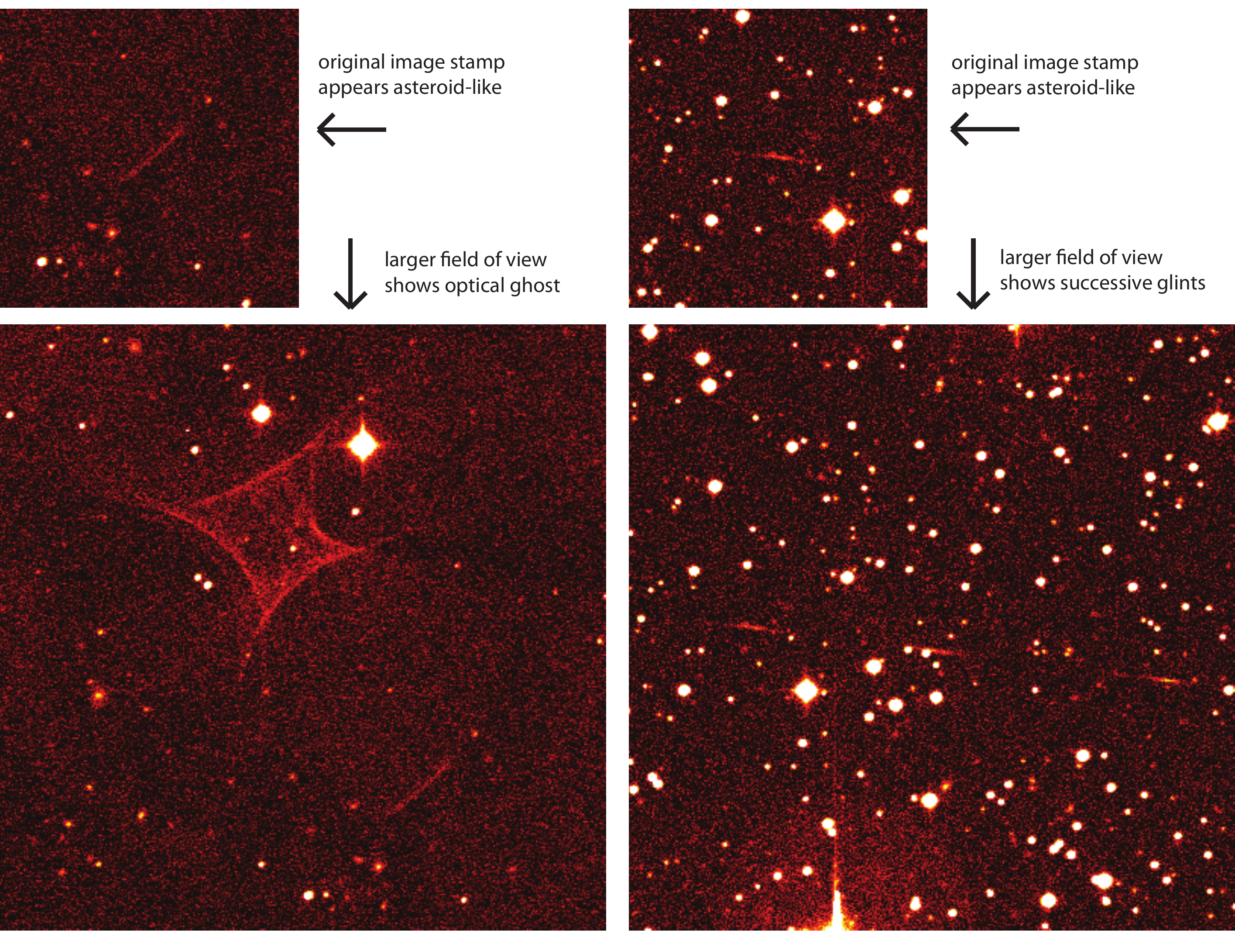}
\caption{Example false positive detections in which the original $200''\times 200''$ image stamp looks like a real asteroid streak, but the larger field of view clearly indicates the nature of the bogus detection. \emph{Left}: Filament of an optical ghost. \emph{Right}: Glint segment, {\it e.g.}, from a fast-moving rapidly-rotating piece of space debris. If additional candidates from these larger false-positive objects also appear on the scanning page, their common exposure timestamp implies their stamps will appear adjacent to one another, facilitating their identification as bogus detections.}
\end{figure*}

In addition to tracking the classifier's performance during the training cross-validation trials, after training we subjected the classifier to a new sample of $\sim$400 synthetics. These newly-generated synthetics were injected into the same 539 test images using the same procedure described in section 3.1.2. Given the distinct random numbers used in this run, these synthetics are distinct from those that were used in training, and appear at different locations on the PTF images.

As was done in cross-validation, the purpose of this post-training trial was to ascertain the detection completeness, though this time using synthetics (which were used previously for training but not testing). Another difference is that we now consider completeness for a fixed classification probability threshold ($p>0.4$) and do so as a function of magnitude and length (similar to the analysis done for \texttt{findStreaks} in Section 3.1.2).

The top plots of Figure~11 show the same information as was shown in Figure~5, albeit for this new sample of synthetics (and at slightly coarser resolution). Namely, we first examine the completeness delivered by \texttt{findStreaks} alone, and again see the limiting magnitude versus length trend. In the bottom plots of Figure~11, we show detection completeness for the same sample only this time for the combined \texttt{findStreaks} plus machine classifier system. In other words, all the blue data points in the lower left Figure~11 plot were both successfully detected by \texttt{findStreaks} \emph{and} were subsequently classified as real with a probability $p>0.4$.

In Figure~12, we again show data from the same synthetics sample, this time plotting the loss in absolute detection completeness due solely to the application of the machine classifier. In the top plot of Figure~12, green data points were successfully detected by \texttt{findStreaks} but did not score high enough ($p>0.4$) in the classification stage. The 2D histogram below it shows that the most significant loss in completeness occurs for short faint streaks. Likely not coincidentally, this region suffers from the largest number of bogus \texttt{findStreaks} detections, as indicated by Figure~6. Integrating over all bins in this magnitude-vs.-length histogram, we observe an average completeness drop of $\sim$0.15, consistent with Figure~10 for a true positive rate of $\sim$85\% accompanying a false-positive rate of $\sim$5\%.

\subsection{Web-based screening interface}

The final component of the discovery portion of the PTF streak-detection pipeline consists of a webpage for human vetting of image stamps of streak candidates to which the classifier has assigned a high probability of being real. Figure~4 includes a screenshot of this webpage. Given the $\sim$5\% false-positive rate quoted in the preceding paragraph and the $\sim$$10^5$ detected candidates accumulated in a typical night (cf. Section 3.2.1), this webpage displays on average several thousand candidates per night.

Including operations on Palomar Mountain, the data transfer from Palomar to Caltech, and the IPAC real-time processing pipeline (Section 2.3) a typical lag-time of $\sim$30 minutes (approx. $\pm$10 minutes) elapses between the acquisition of exposures with the PTF camera and the posting of streak candidates from said exposure to the scanning webpage. The image stamps have fields of view of $200''\times 200''$ with linear contrast scaling from $-$0.5$\sigma$ to 7$\sigma$ (as in Figure~3). Undifferenced images are reviewed as opposed to the differenced images, to better provide context to the scanner and enable him/her to visually assess the observing conditions ({\it i.e.} the density and image quality of background stars).

\begin{table*}
{\footnotesize
\caption{PTF discoveries of streaked NEAs (between 2014-May-01 and 2014-Dec-01)}
\begin{tabular}{lcccccccccc}
\hline
\multirow{2}{*}{name} & \multirow{2}{*}{date} &   \# PTF  &diameter (m)& inclination &\multirow{2}{*}{eccentricity} & semi-major & min. Earth orbit & $\Delta v$ w.r.t.& speed      & $V$  \\
                      &                       &detections&(7\% albedo)& (deg)      &                      & axis (AU)                 & intersect. (LD)   & Earth (km/s)     & ($''$/min) & magnitude    \\
\hline
2014 WS$_7$           &Nov-19   & 3   & 17.4     &  8.687     &  0.47655            &  1.8918           & 4.36                 & 6.1              & 25.7           & 18.7       \\
2014 WK$_7$           &Nov-18   & 4   & 166      &  23.91     &  0.36071            &  1.5605           & 3.38                 & 8.4              & 39.8           & 16.7       \\
2014 UL$_{191}$       &Oct-30   & 2   & 66.2     &  2.070     &  0.65017            &  1.7553           & 3.73                 & 7.8              & 66.9           & 17.1       \\
2014 ST$_{223}$       &Sep-23   & 4   & 15.2     &  5.875     &  0.17045            &  1.0532           & 1.56                 & 5.2              & 31.8           & 18.4       \\
2014 SE$_{145}$ &Sep-23\footnote{Designated by MPC as ``First observed at Palomar Mountain--PTF on 2014-09-23'', but MPC also cites Pan-STARRS 1 observations on Sep-22}&  2   & 18.6     & 8.406     & 0.54698		 &  2.2354		& 4.03		& 6.6			& 62.3	& 19.1 \\
2014 SC$_{145}$       &Sep-23   & 3   & 36.4     &  20.17     &  0.20651            &  1.2216           & 6.61                 & 7.8              & 28.4           & 19.0       \\
2014 SE               &Sep-16   & 2   & 43.4     &  20.02     &  0.17187            &  1.2428           & 13.5                 & 7.9              & 24.7           & 18.7       \\
2014 RJ               &Sep-02   & 2   & 41.8     &  19.57     &  0.27083            &  1.4086           & 7.39                 & 7.4              & 37.2           & 17.9       \\
2014 LL$_{26}$        &Jun-09   & 2   & 43.7     &  9.182     &  0.10177            &  1.1427           & 5.25                 & 5.4              & 29.1           & 17.2       \\
2014 KD               &May-17   & 4   & 66.2     &  5.238     &  0.54605            &  2.1519           & 7.74                 & 6.2              & 30.0           & 17.5       \\
2014 JG$_{55}$        &May-10   & 4   & 7.26     &  8.739     &  0.41257            &  1.5843           & 0.336                & 5.8              & 60.0           & 18.1       \\
\hline
\end{tabular}
\bigskip
}
\end{table*}

The kinds of false positives commonly encountered on the scanning webpage include all of those shown on the right-hand side of Figure~7. Image stamps are viewed in chronological order, so that candidates from a common image appear consecutively on the scanning page. This enables rapid recognition of false positives of a common origin. For example, multiple segments of a long satellite trail, large optical ghost, or artifacts from a poorly-subtracted or high stellar density image will appear together and are thus easily dismissed. Artifacts that do not appear in groups, such as cosmic ray hits, background sky noise and poorly-subtracted galaxies, are rapidly visually dismissed as well. A full night's set of candidates (several thousand) can be reliably reviewed by a trained scanner in 5--10 minutes, though the reviewing time is distributed over the during observing session, as the webpage is refreshed every 20--30 minutes.

Clicking on the image stamp of a candidate streak presents another webpage with more detailed information including astrometry, photometry, \texttt{realBogus} score, image stamps of the differenced and reference images, and a larger field of view around the detection. Certain types of false positives are more easily identified using this additional information, including portions of optical ghosts and periodically-glinting space debris. This summary page also contains information for real-time follow-up, as discussed in the next section.

\section{Follow-up and reporting of discoveries}

Once a real streak is discovered in PTF via the steps outlined in the previous section, we trigger real-time follow-up with the same telescope. Its wide field of view (Section 2.1) makes the PTF camera particularly well-suited for recovering fast-moving NEAs within a few hours of an initial detection. As described below, the follow-up process effectively interrupts the nominal robotic survey by injecting high-priority exposure requests into the queue. The final step involves reporting observations to the Minor Planet Center to facilitate subsequent confirmation and follow-up worldwide.

\subsection{Target-of-opportunity (ToO) requests}
\label{sec:too}

The sequence of PTF fields observed on any given night is determined in real-time by a robotic scheduler: the P48 Observatory Control System (OCS) described by \cite{law09}. The robot takes as input a list of fields, generally prescribed by a human operator per lunation, and attempts to optimize exposure conditions (distance from moon, airmass, etc.) while also maintaining a specific cadence---predominantly two or three exposures per field per night separated by $\sim$40 minutes (optimal for supernova discovery). As noted in Section 2.1, in recent years (during the iPTF phase), fields and cadences have often been allocated to distinct experiments, though all exposures still adhere to a fixed tiling of fields, with 60-second integrations in either $R$- or $g$-band.

All PTF exposures are processed by the streak detection pipeline if they have a reference image available (required for image differencing, see Section 2.3). Upon recognition of a single detection of a likely real NEA streak on the scanning webpage, the human reviewer immediately checks the webpage for additional serendipitous detections in other PTF exposures acquired that night. 
Given the small number of same-night candidates, the human reviewer checks whether the direction and angular velocity of the same-night streaks are consistent with the measured RA and DEC of the two streaks. If a second detection is determined to be from the same object, 
the observations are immediately sent to the MPC's Near-Earth Object Confirmation Page (NEOCP)\footnote{\href{http://www.minorplanetcenter.net/iau/NEO/toconfirm_tabular.html}{\color{blue}{http://www.minorplanetcenter.net/iau/NEO/toconfirm\_tabular.html}}}.
The scanning webpage assists in the matching process by showing, on each candidate streak detection's detailed summary page, all fields of view acquired the same night centered on extrapolated positions of the object assuming constant-velocity great-circle motion (in both directions). If a second detection is visually found, the human scanner may then search the webpage's list of positive detections at the specific timestamp/field corresponding to the visually-confirmed second detection.

Lacking a second detection, the reviewer uses tools integrated into the scanning webpage to trigger \emph{target-of-opportunity} (\emph{ToO}) exposures to secure additional detections. Figure~4 shows a screenshot of the webpage's streak position estimation tool, which uses a linear (great-circle) extrapolation assuming motion in either direction, overlaid on the PTF tile grid. A PHP script redraws the plot to the current time when refreshed by the user.

Once a list of fields potentially containing the streak has been identified (typically between one and a few fields), a text-based email sent to the telescope robot inserts the fields into the queue with very high weight. This email may additionally prescribe repeat exposures of the fields with some specified cadence, filter, or maximum airmass. The ToO exposures typically are acquired within 5--10 minutes of the request, depending on factors such as slew time and the need to change filters. The email-based ToO-system for PTF was originally designed for (and proven on) the discovery of optical afterglows of gamma-ray bursts \citep{sin13}.

Apart from having been manually triggered, the ToO exposures are otherwise identical to routine PTF survey images in that they are sidereally tracked, 60-second $R$- or $g$-band images aligned to a fixed tile grid of the sky (as opposed to, {\it e.g.}, being centered on the NEA's predicted position). Having acquired the ToO exposures, any additional detections of the streak are automatically extracted with the same streak-detection pipeline and will appear on the scanning webpage along with the rest of the night's candidates. Observations are sent to the MPC once two or more detections have been secured.
The information sent to the MPC includes an estimate of the center point of the streak as well as the time of the mid-point of the observation. 
The accuracy of shutter opening and closing is about $\pm 10$~ms.  Once two streak detections are made, the order of the endpoints is determined and in principle
an astrometric estimate of the locations of the endpoints could be provided to the MPC, together with their respective times.  This will
be considered for a future submissions to the MPC.

\begin{figure}
\centering
\includegraphics[scale=1.11]{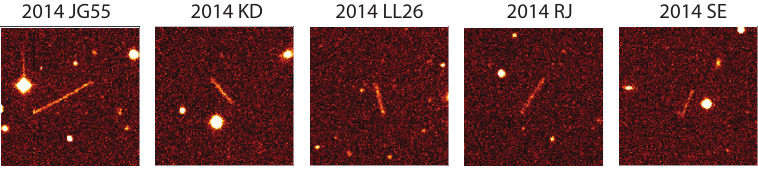}
\caption{Discovery images of the first five streaked NEAs found by PTF.}
\end{figure}

\begin{table}
\small{
\centering
\caption{iPTF sub-surveys containing streaked-NEA discovery exposures.}
\begin{tabular}{lcccccccccc}
\hline
\multirow{2}{*}{name}  & sub-survey in which  & \multirow{2}{*}{filter}  & degrees from\\
                       & NEA was discovered           &                          & opposition \\
\hline
2014 WS$_7$           &Permanent Local Galaxies & $R$  & 28\\
2014 WK$_7$           &TILU K2 Campaign         & $g$  & 73\\
2014 UL$_{191}$       &TILU Fall 2014           & $g$  & 47\\
2014 ST$_{223}$       &Opposition NEA search    & $g$  & 18\\
2014 SC$_{145}$       &RR Lyrae                 & $R$  & 32\\
2014 SE               &RR Lyrae                 & $R$  & 21\\
2014 RJ               &TILU Fall 2014           & $g$  & 35\\
2014 LL$_{26}$        &Star-forming low-cadence & $R$  & 9\\
2014 KD               &TILU Spring 2014         & $R$  & 49\\
2014 JG$_{55}$        &iPTF14yb follow-up       & $R$  & 34\\
\hline
\end{tabular}
\bigskip
}
\end{table}

\begin{figure*}
\centering
\includegraphics[scale=0.75]{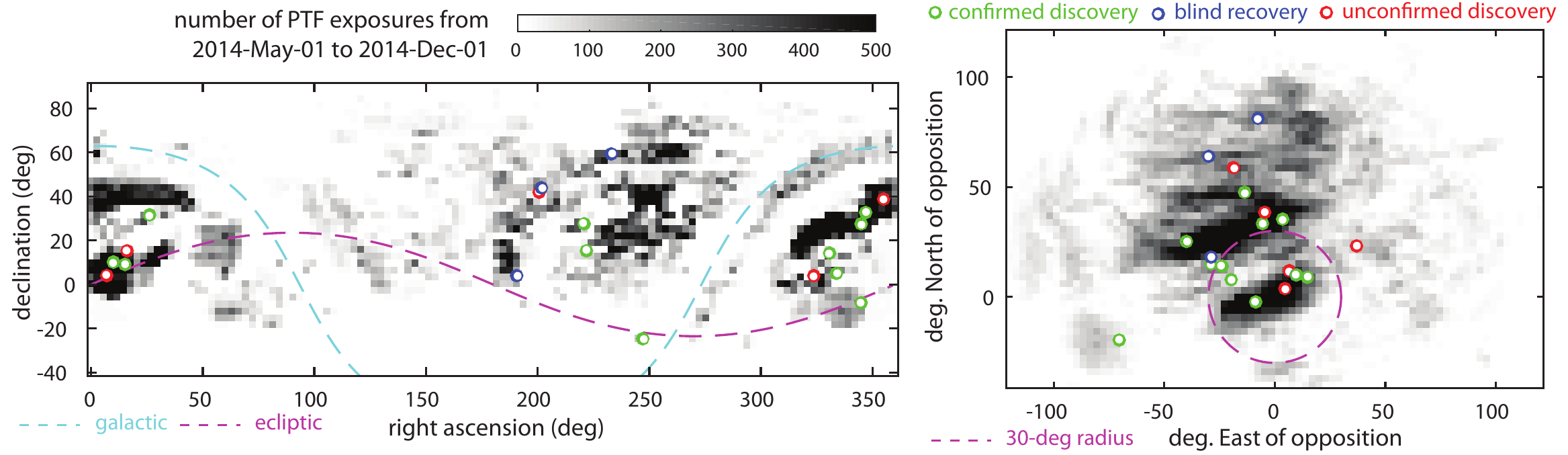}
\caption{Distribution of PTF exposures (\emph{left}: in sky coordinates, \emph{right}: with respect to opposition) and streaked NEA detections (\emph{right}: with respect to opposition) from 2014-May-01 through 2014-Dec-01. The grayscale scalebar maps the density of PTF exposures in both plots. Exposures for which realtime streak-detection was not performed are not included ({\it e.g.}, fields lacking reference images or with too high source density on the galactic plane). }
\label{fig:ptf_exposures}
\smallskip
\end{figure*}

\subsection{Initial NEA discoveries}

The full streak-discovery system, incorporating the IPAC real-time data products, \texttt{findStreaks} and the trained machine classifier, began real-time operations 2014-May-01. About a week later, the first PTF streaking NEA discovery was made (2014 JG$_{55}$). Passing at one-third of a lunar distance, this object is also the smallest and closest-approaching NEA yet discovered by PTF.

The largest streaking NEA discovered by PTF to date is 2014 WK$_7$, at $H=22.4$ mag ($D\approx 166$ m), while the PTF discovery having an orbit with the lowest $\Delta v$ with respect to Earth and most accessible by a robotic space mission (e.g. see \citealp{jed13} ) is 2014 ST$_{223}$.

Table~2 details the eleven total streaking-NEA discoveries made by PTF as of 2014-Dec-01. Nearly all of these (the one exception being 2014 LL$_{26}$) were followed up and confirmed by multiple observatories within 24 hours. A total of 25 different observatories have provided follow-up observations within 24 hours of at least one of the NEA discoveries listed in Table~2. After the sun has risen in California, most short-term follow-up of PTF discoveries occurs from Japan and Europe (occasionally Australia), as most longitudes west of Palomar fall in the Pacific.

Figure~15 shows the discovery position of the NEAs in Table~1 relative to opposition. 
A majority of the objects were found within 40$^\circ$ of opposition. 
An outlier is 2014 WK$_7$, which was discovered 73$^\circ$ from opposition (phase angle 71$^\circ$), though this NEA is also an outlier in the sample in terms of its size.

Table~3 lists the various sub-surveys (also known as `iPTF experiments', see Section 2.1) to which the NEA discovery exposures belong. Here `TILU' stands for Transients in the Local Universe'. A key point here is that nearly all of PTF's streaked NEA discoveries to date have been made in images originally purposed for non-solar-system science. A dedicated iPTF experiment designed to maximize the area covered around opposition was carried out for several nights in Fall 2014, though only one exposure from said program produced a discovery, 2014 ST$_{223}$.

All follow-up was unsolicited apart from having posted the discoveries on the NEOCP, and attests to the dedication of the worldwide NEA follow-up community. We note however that, while they are on the NEOCP, PTF-discovered streaking NEAs are consistently the \emph{brightest} on the list---all were $V\le 19$ mag---whereas most of the 50+ objects typically found on the NEOCP have $V\ge 20$ mag. It is therefore not surprising that more follow-up facilities are able and willing to recover these bright objects as compared to the typical faint and slow NEOCP candidates.

\subsection{Blind real-time recovery of known NEAs}

There are several options for querying a given R.A., Dec., and time to search for a match (within some radius) to an asteroid with a known orbit; these include MPChecker\footnote{\href{http://www.minorplanetcenter.net/cgi-bin/checkmp.cgi}{\color{blue}{http://www.minorplanetcenter.net/cgi-bin/checkmp.cgi}}}, JPL's HORIZONS\footnote{\href{http://ssd.jpl.nasa.gov/sbfind.cgi}{\color{blue}{http://ssd.jpl.nasa.gov/sbfind.cgi}}}, and PyMPChecker \citep{kle09}. However, those scanning the PTF streak candidates in real-time are discouraged from checking if a detected streak is a known object prior to obtaining ToO follow-up and submitting the observations to the NEOCP. One reason is that the above mentioned query tools are not necessarily reliable for fast-moving objects, and will not always return a match even if the object has a well-determined orbit. Another reason is that the ToO-submitting procedure, while simple and straightforward, requires efficiency and efficacy on the part of the scanner and so should be practiced as often as possible.
Lastly, the MPC encourages submission of unidentified known objects as it allows them to directly assess our program's detection capabilities ({\it e.g.}, our astrometric accuracy).

As of 2014-Dec-01, a total of three previously-discovered NEAs have been blindly detected by PTF as streaks and submitted to the NEOCP: 2014 HL$_{129}$ (May-02); 2010 JO$_{33}$ (May-08); and 2014 WF$_{108}$ (May-27; to date the only `potentially hazardous asteroid' blindly detected as a streak by PTF in real-time).

\subsection{Unconfirmed discoveries}

A total of five PTF objects posted to the NEOCP (between May-01 and Dec-01) did \emph{not} receive external follow-up, meaning they never obtained confident orbit solutions and thus were not assigned provisional designations by the MPC (Table~4 and Figure~16). For four of these unconfirmed objects, PTF had submitted only two observations to the NEOCP. We note that 5 out of the 11 \emph{confirmed} objects (Table~2) also were reported with only two observations, from which we naively conclude that a two-observation discovery has only a 50\% probability of being successfully followed-up (for three- and four-observation discoveries the recovered fraction increases to 66\% and 100\%, respectively).

While in reality the recovery probability depends also on the temporal spacing of the observations, the object's speed and magnitude, and the availability of follow-up resources ({\it e.g.}, less facilities operate around full-moon), the number of observations seems to be a useful indicator of the recovery likelihood. Users of the PTF real-time scanning and ToO system attempt to obtain at least three observations for discoveries, though this is not always possible, {\it e.g.}, for discoveries made early in the night in the western sky, or just before dawn. Occasional technical issues with the real-time processing and/or ToO system also can hinder PTF self-follow-up.

\subsection{Artificial satellites}

\begin{table}
\small{
\centering
\caption{Unconfirmed PTF streak discoveries (from 2014-May-01 to 2014-Dec-01)}
\begin{tabular}{lcccccccccc}
\hline
NEOCP         & date & num. & speed       &  $V$    & notes   \\
name          & found& obs. & ($''$/min)  &  (mag)  &       \\
\hline
PTF5i5        & May-04 &  2    &   46.8           &  19.5     & \\
PTF9i2        & Jul-08 &  2    &   36.9           &  17.9     & 85\% moon, near dawn\\
PTF3k8        & Sep-23 &  2    &   64.9           &  18.0     & likely satellite\\
PTF8k2        & Sep-25 &  3    &   27.6           &  18.9     & \\
PTF7l3        & Oct-25 &  2    &   30.1           &  17.5     & \\
\hline
\end{tabular}
}
\end{table}

\begin{figure}
\centering
\includegraphics[scale=0.44]{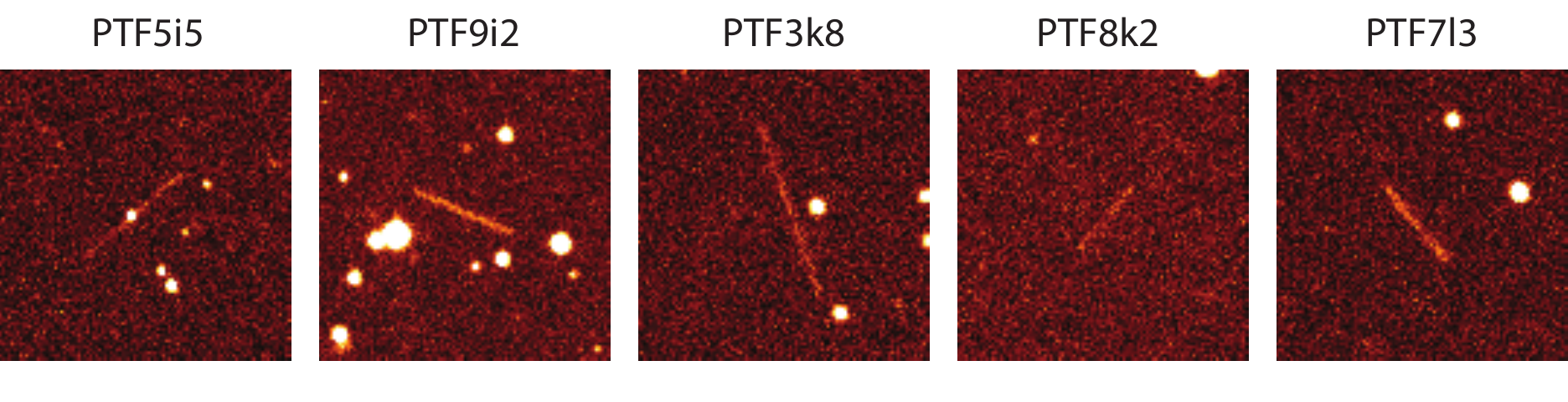}
\caption{PTF streak discoveries that were posted to the NEOCP but never received external follow-up.}
\smallskip
\end{figure}

\begin{figure}
\centering
\includegraphics[scale=0.36]{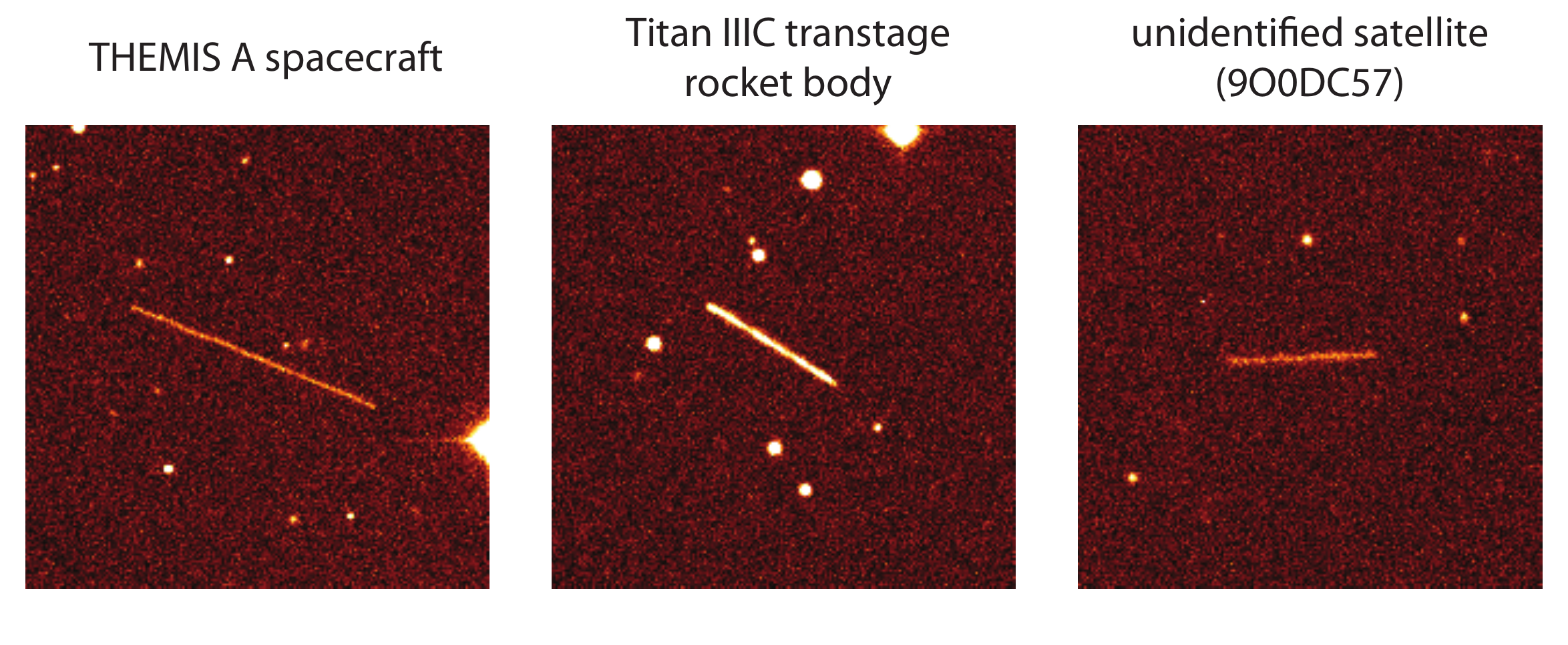}
\caption{Artificial satellites detected as streaks by PTF (identifications provided by the MPC).}
\end{figure}

Many distant Earth-orbiting artificial satellites can, at certain parts of their orbit, appear consistent with an Earth-approaching NEA. Our streak-recognition pipeline has on several occasions detected such satellites. Figure~17 shows some examples, including one of the THEMIS mission spacecraft studying the Earth's magnetosphere \citep{ang08} and a Titan IIIC rocket body. The MPC's automated observation-ingestion processes outputs known artificial satellite matches to NEOCP submissions (as was the case for the three in Figure~17), though in some cases the object will be posted to the NEOCP and remain on the list for some time prior to its recognition as artificial. Three examples of the latter were PTF7i2, PTF8i6, and PTF0n2.

While we see the same value in blind reporting of artificial satellites as we do blind reporting of known NEAs (Section 4.3), some high-orbit satellites have geosynchronous orbits and can therefore appear in the same area of sky for many consecutive nights. An example is the THEMIS spacecraft, whose apogee was coincident with opposition, causing it to be repeatedly observed by PTF in autumn 2014. For routine identification of known satellites, we have therefore adopted the useful software tool \texttt{sat\_id} by Project Pluto\footnote{\href{http://www.projectpluto.com/sat_id.htm}{\color{blue}{http://www.projectpluto.com/sat\_id.htm}}}.

\section{De-biased detection rate}

\begin{figure}
\centering
\includegraphics[scale=0.75]{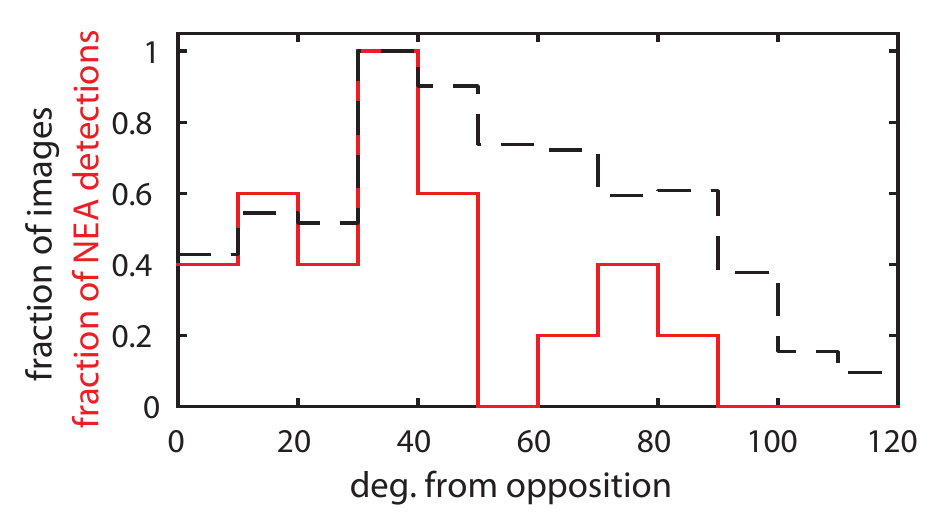}
\caption{Normalized distributions of PTF images and streaked NEA detections with respect to opposition. The 19 NEAs included here consist of new discoveries, blind recoveries and the five unconfirmed discoveries. See Figure~15 for the two-dimensional distribution.}
\end{figure}

\begin{figure}
\centering
\includegraphics[scale=0.75]{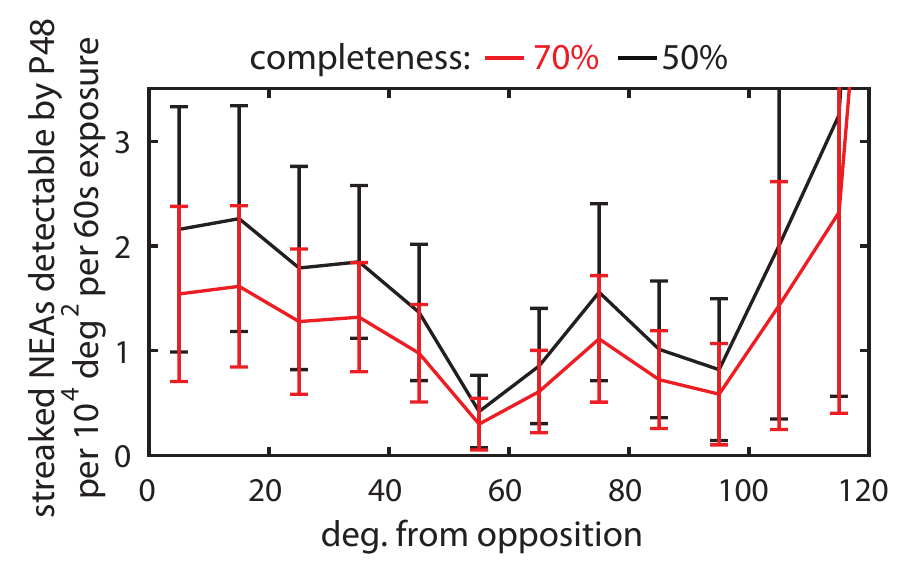}
\caption{Estimates of the number of streaked NEAs detectable by P48 as a function of distance from opposition. Computed using the data in Figure~18 and Equation (5).}
\end{figure}

The right panel of Figure~15 shows the distribution of streaked NEA detections (including confirmed and unconfirmed discoveries as well blind recoveries). In this section we use this sample of detections and the distribution of PTF exposures with respect to opposition to derive the de-biased streaked-NEA detection rate as a function of radial distance from opposition. Figure~18 shows the same data as in Figure~15, removing the azimuthal information to only show the one-dimensional radial distributions.

We seek to estimate the frequency $f$ of streaked NEA detections per unit area of sky per unit time (equivalently, per survey image). The posterior probability distribution of $f$ (assuming a constant prior) is given by an appropriately-normalized Poisson distribution:

\begin{equation}
P(f)=\frac{(NC)^{n+1}}{\Gamma(n+1)}f^n \exp(-NCf)
\end{equation}

\noindent where $N$ is the total number of images searched for streaked NEAs, $n$ is the number of detected streaked NEAs, $C$ is the completeness (true positive rate) of the PTF streak detection system as a whole, and $\Gamma(\ldots)$ is the gamma function (which contributes to the normalization of the distribution).

Figures 5 and 11 indicate that the completeness $C$ depends on which volume in magnitude, length, and orientation space under consideration, as well as the separate efficincies of sub-components like \texttt{findStreaks} and the machine classifier. For simplicity, in the following analysis we evaluate two separate values for $C$ (0.5 and 0.7) but the most accurate values for $C$ would in principle come from direct application of the completeness data in Figures 5, 11 and 12.

We apply Equation (5) to the image count $N$ and streak count $n$ within each of the thirteen bins in Figure~18. 
In particular, by numerical integration we compute the 16th and 84th percentiles of the resulting Poisson distributions, and plot these 
bounds as a function of distance from opposition in Figure~19. The estimates are 1--3 streaked NEA detections per 10$^4$ deg$^2$ of sky \emph{near opposition}, dropping to about 1 or less beyond 40--50 deg from opposition. The images acquired by PTF from 2014-May-01 through 2014-Dec-01 represent 191,435 deg$^2$, and a total of 19 streaked NEAs (10 confirmed, 4 blindly recovered, 5 unconfirmed) were detected in these data. If the areal density of the streaks were independent of distance from opposition, this would correspond to a coarse estimate of $\sim$1 detected streak per 10$^4$ deg$^2$, in agreement with the radially-binned rates multiplied by the actualy radial distribution of images (which are mostly 40 deg or more from opposition).

\section{Scaling laws for streaked asteroid detection}
\label{sec:scaling}

We here derive a quantitative `figure of merit' (FoM) proportional to the average number of streaked asteroids detectable per unit time by a survey. 
The FoM will depend on a number of survey specifications including the:

\begin{itemize}
\item field of view $\Omega$ in deg$^2$,
\item seeing width $\theta_\text{PSF}$ in arcseconds,
\item limiting magnitude of a point source $m_\text{lim}$, for a given exposure time $\tau$, and
\item the time between exposures $\tau_\text{tot}$, which includes readout and telescope slew time.
\end{itemize}

Assume that the density of asteroids and their velocity distribution is independent of distance. The volume of streaked asteroids detectable at any given time goes as $\Omega D_\text{strk}^3$, where $D_\text{strk}$ is the maximum distance at which an asteroid can be detected as a streak. The figure of merit (asteroids detectable per unit time) therefore scales as

\begin{equation}
\text{FoM} \propto \frac{\Omega D_\text{strk}^3}{\tau_\text{tot}}
\end{equation}

For a point source, $D_\text{pnt}^2 \propto 10^{0.4 (m_\text{lim}-m_0)}$, where $m_0$ is a normalization and consequently we define
the point source FoM to be

\begin{equation}
\text{FoM}_\text{pnt} \equiv \frac{\Omega 10^{0.6 (m_\text{lim}-m_0)}}{\tau_\text{tot}}
\end{equation}
\noindent
A principal result, shown in the appendix, is that the FoM for long streaks is simply

\begin{equation}
\text{FoM}_\text{strk} = \text{FoM}_\text{pnt} \left(  \frac{\theta_\text{PSF}}{\theta_\text{strk}} \right)
\end{equation}
\noindent
where $\theta_\text{strk}$ is the angular length of the asteroid streak for an asteroid with a given
velocity perpendicular to the line of sight at the limiting point source detection distance for the asteroid.

Streak detection thus differs significantly from the more conventional
analysis that was designed for detection of asteroids with
point-like appearance in individual images.
It is therefore of interest to estimate the capabilities of various
current and future surveys assuming that those surveys implemented a streak detection
capability of the type described in this paper. 

For deep surveys with larger aperture telescopes (in particular DECam, SST, and LSST), many
small NEAs will be detected as point sources because they can be detected
far from earth where the apparent angular velocity will be small.
Consequently, for these surveys, streak detection will provide a significant improvement
only for the smallest fast-moving asteroids.
For shallower surveys, such as PTF and ZTF, streak detection can significantly
improve the capability for discovery of small NEAs.

In the appendix, we derive an analytical expression for  FoM,
proportional to the average number of streaked asteroids detectable per unit time.
As defined, the FoM is proportional to the volume per unit time surveyed by a  telescope
for an asteroid of given normalized brightness and given velocity (direction and magnitude).
We discuss the analytical form of a generalized FoM that is applicable to
both point source 
detection and streaked asteroid detection and
transitions smoothly between the two regimes for any survey. 
This computation can be done for any desired representative class of small
NEAs,  specified by their brightness ($H$ magnitude) and asteroid velocity perpendicular
to the line of sight ($v_\perp$).
We also discuss the numerical integration of the generalized FoM over an isotropic distribution
of angles to obtain an FoM for asteroids of a given $H$ magnitude and velocity relative to earth ($v_\text{ast}$),
taking into account the relative numbers of objects with a given perpendicular velocity
to the line of sight.

\begin{table*}[ht] 
\caption{Comparison of relative figure-of-merit for detection of small near earth asteroids. 
The H magnitudes correspond to 10~m, 50~m, 100~m, and 200~m asteroids (for H=28.5, 25.0, 23.5, and 22.0, respectively, assuming an albedo of 0.07 ).
The two FoMs in each case are with and without implementation of streak detection, the latter case (peak detection) is in parenthesis.
Variables are further defined in the text and the appendix. The FoMs are averaged over an isotropic distribution of
velocities of magnitude, $v$, computing the asteroid velocity, $v_\perp$, perpendicular to the line of sight for each direction.}
{
\centering  
\vspace{0.5ex}
\begin{tabular}{|cccccc|c|c|c|c|c|c|c|c|}  
\hline                    
Telescope&$\Omega$&$m_{lim}$&$\theta_\text{PSF}$&$\tau$&$\tau_{tot}$&\multicolumn{2}{|c|}{$\left< \text{FoM} \right>$}&\multicolumn{2}{|c|}{$\left< \text{FoM} \right>$}&\multicolumn{2}{|c|}{$\left< \text{FoM} \right>$}&\multicolumn{2}{|c|}{$\left< \text{FoM} \right>$}\\
\footnotesize
&(deg$^2$)&(mag)&(arcsec)&(s)&(s)&\multicolumn{2}{|c|}{(H=28.5)}&\multicolumn{2}{|c|}{(H=25.0)}&\multicolumn{2}{|c|}{(H=23.5)}&\multicolumn{2}{|c|}{(H=22.0)}\\
&& &  & &&\multicolumn{2}{|c|}{(v=20~km/s)}&\multicolumn{2}{|c|}{(v=20~km/s)}&\multicolumn{2}{|c|}{(v=20~km/s)}&\multicolumn{2}{|c|}{(v=5~km/s)}\\ [0.5ex]
\hline\hline
PTF & 7.25 & 20.7 & 2.0 & 60 & 105 &\multicolumn{2}{|c|}{$\equiv$1} &\multicolumn{2}{|c|}{$\equiv$1}&\multicolumn{2}{|c|}{$\equiv$1}&\multicolumn{2}{|c|}{$\equiv$1}\\[0.5ex]
\hline
&&&&&&{\it Streak} & {\it Peak}&{\it Streak} & {\it Peak}&{\it Streak} & {\it Peak}&{\it Streak} & {\it Peak}\\
\hline
DECam & 3.2$^a$ & 23 &0.8 & 40 & 60 &31&(1.4)&28&(6)&26&(11)&20&(18)\\
PS1& 7 & 21.8 & 1.1 & 30 & 40 &20&(1.0)&18&(4.4)&17&(7)&13&(11)\\
ZTF & 47 & 20.4 & 2 & 30 & 45 &17&(0.8)&15&(3.5)&14&(6)&11&(9)\\   
ATLAS & 60 & 19.9 & 2.5 & 30 & 35 &14&(0.6)&13&(2.8)&12&(5)&9&(8)\\
CSS/MLS-II&5.&21.5&1.3&20&35&17&(1.2)&14&(5)&12&(6)&7&(6)\\
CSS-II & 19.4 & 19.5$^b$ & 1.5 & 30 & 45 &1.0&(0.02)&1.1&(0.1)&1.1&(0.2)&1.2&(0.9)\\[0.4ex]
\hline
SST/Lincoln&6.&20.5&1.&2.&6.&96&(33)&44&(33)&29&(25)&14&(13)\\[0.4ex]
LSST & 9.6 & 24.4 &0.7 & 30 & 39 &2000&(200)&1500 &(650)&1200&(700)&700&(650)\\[0.4ex]
\hline    
\end{tabular} 
}
\begin{spacing}{0.75}
{\footnotesize {\bf General notes}: $\Omega$ is the field of view.  $m_\text{lim}$ is the 5$\sigma$ median limiting r-band magnitude. $\theta_\text{PSF}$ is the width of the telescope point spread function.
$\tau$ is the exposure time and $\tau_\text{tot}$ is the time between exposures.\newline 
{\bf PTF}: $m_{lim}$ is the measured PTF value. \newline
{\bf DECam}: $\Omega$, $m_\text{lim}$ and $\tau$ from \cite{allen+15}. Note $^a$: the value of $\Omega$ quoted by \cite{allen+15} for NEO observations is less than the 3.2 deg$^2$ normally quoted.  $\theta_\text{PSF}$ from \cite{shaw2015noao}, Table~4.1. $\tau_{tot}$  assumes a read time of 20s (from  \cite{shaw2015noao}, Table~4.2). \newline
{\bf PS1}:  Performance is from the 3$\pi$ survey of PS1. Recent upgrades may have improved performance for PS1/2 \citep{morgan+12}. $\Omega$  from \cite{kaiser04}. $m_{lim}$ is median r-band
limiting magnitude from \cite{morganson+12}. $\theta_\text{PSF}$ from \cite{lee+12}.\newline
{\bf ZTF}:  $m_\text{lim}$ is the 5$\sigma$ median limiting r-band magnitude estimated using PTF observing history and accounting for new optics and shorter exposure time.\newline
{\bf ATLAS}: The dark sky (not median) magnitude limit in r-band is 20.2 (from \cite{atlas2016}). We take $m_\text{lim}\sim 19.9$ as an estimate of an achievable median performance. Other values from \cite{atlas2016}.\newline
{\bf CSS/MLS-II (Mt. Lemmon upgrade)}: $\Omega$ = 5 sq deg from \cite{christensen+15}. $\theta_\text{PSF}$ from \cite{jedicke+16}. Other parameters from \cite{lar07} except for
$\tau_\text{tot}$ which is estimated.\newline
{\bf CSS-II (Catalina upgrade)}: $\Omega$ = 19.4 sq deg, \cite{christensen+15}. Other values from \cite{mahabal16}.  Note $^b$: 
CSS-II may be able to detect NEOs at less than 5$\sigma$, perhaps as low as 1.2$\sigma$ as described in \cite{christensen14} for CSS. See discussion in text. \newline
{\bf SST/Lincoln}: $\Omega$, $\tau$, and $\tau_\text{tot}$ from \cite{ru14}. $\theta_\text{PSF}$ is estimated, including seeing. $m_\text{lim}$ from \cite{ruprecht+15}.\newline
{\bf LSST}: $\Omega$, $\tau$, and $\tau_\text{tot}$ from \cite{lsst2016}. $m_\text{lim}$ and $\theta_\text{PSF}$ from \cite{ivezic+08}. \newline
}
\end{spacing}
\label{tab:surveys}  
\end{table*}

Table~\ref{tab:surveys} summarizes the results for various surveys.
The choice of surveys is selective, with emphasis on current and near-future capabilities.
PTF with streak detection is used as the reference in all cases.
Two FoMs are given for each survey and each case of brightness and velocity of the NEA.
The first number in each case is the FoM for streak detection, relative to the streak detection FoM for PTF.
The second number (in parentheses) is the FoM for ``peak'' detection relative to the streak detection FoM for PTF.
The FoM for peak detection is described in the appendix and corresponds to the MOPS approach used by
many asteroid surveys, namely, identification of a candidate asteroid when the flux integrated over the typical point spread function is above a
given threshold.

The estimates of relative FoM are only approximate because of the uncertainty of the values for median
performance for each survey. 
For instance, an error of 0.15 magnitudes in the median limiting magnitude, $m_{lim}$, will appear as a 25\% error in the FoM.
We also note that the FoM is a detection volume and the number of NEAs actually detected in any observation
depends critically on many factors: on the direction in which the observation is made, e.g. in the ecliptic, at opposition, etc., 
the cadence of repeat observations of a given field,
as well as other observational parameters such as phase of moon, airmass, etc.
Assessment of these factors for a survey typically requires a detailed simulation.
The FoM is attempts
to quantify an instrument's potential for carrying out a small asteroid survey.
How a given instrument is used for asteroid observations will ultimately determine its effectiveness.
Individual surveys obviously differ significantly in their overall survey strategy. 
For example, for the PTF observations reported here, the overall survey strategy was largely determined by objectives other than solar system
science, such as supernova science, resulting in about 2/3 of the observations being greater than 20 degrees from the 
ecliptic plane (see e.g. Figure \ref{fig:ptf_exposures}).

The H magnitudes shown in the table, 28.5, 25.0,23.5, and 22.0 correspond approximately to NEAs of 10~m, 50~m, 100~m, and 200~m
in size, for an albedo $\sim 0.07$.  
A velocity of 20~${\rm km/s}$ is chosen as typical (\cite{jeffers+01}).
The circumstances chosen span a range of size, with the longest streaks expected for the case
of $H=28.5$ and $v_\text{ast}=20$ km/s and the most point-like objects to be seen in the case of $H=22.0$ and $v_\text{ast}=5~{\rm km/s}$.
Table~\ref{tab:surveys} shows that for most of the telescopes streak detection is about 20 times more efficient than conventional peak detection for
very small fast moving asteroids ($H=28.5$, $v_\text{ast}=20~{\rm km/s}$), 
about 4 times more efficient for 50~m asteroids ($H=25.0, v_\text{ast}=20~{\rm km/s}$), 
about 2.5 times more efficient for 100~m asteroids ($H=23.5$, $v_\text{ast}=20~{\rm km/s}$), and, as expected, only slightly
more efficient at detecting slower-moving 200~m asteroids ($H=22.0$, $v_\text{ast}=5~{\rm km/s}$).
The FoM values shown for $H=22.0$, $v_\text{ast}=5~{\rm km/s}$ are in fact reasonably well approximated by the
simple point source FoM given in Eq. \ref{eq:fom_pnt} in the Appendix.

Taking into account that the estimates of FoM in Table~\ref{tab:surveys} are only approximate, we can make some
general statements.
For the four cases of asteroid size and velocity  shown in Table~\ref{tab:surveys}, 
the capabilities of DECam, PanSTARRS, ZTF, ATLAS, and CSS/MLS-II are similar in terms of performance per unit time if implementing a similar survey strategy. 
The FoM values for these 5 surveys are within about a factor two of each other and therefore 
the total time for which asteroid surveys are carried out
and the choice of survey pointing strategy will be the significant factors in determining the
relative long-term productivity of each of the surveys for small NEO detection.
Clearly, both SST/Lincoln and LSST represent a major improvement in capability for small asteroid detection.

To a large extent, the PTF observations described in this paper were directed towards the detection of $\sim$10~m asteroids.
Although PTF is nominally about 13 times less effective than PanSTARRS for detecting slow moving 200~m NEAs
(entry for H=22. for PS1 in Table~\ref{tab:surveys}), using streak detection it has roughly the same effectiveness per unit time as
PanSTARRS for fast moving 10~m asteroids; our estimate is that PanSTARRS would be $\sim$ 20 times more effective than PTF if 
streak detection were implemented (entry for H=28.5).  
This is borne out by the statistics of small NEA detection for the period during which PTF carried out its small NEA demonstration,
1 May 2014 to 1 December 2014. 
The statistics were taken from MPC data on discovered NEOs for that period.
During this period PTF discovered one NEA with H$>28$, 2014 JG55, while PanSTARRS-1 detected two such NEAs.
For H$>27$, PTF discovered four NEAs, while PanSTARRS discovered nine.
This indicates that PTF was roughly half the effectiveness of  PanSTARRS in discovering the smallest NEAs,
This is consistent with the predictions of Table~\ref{tab:surveys}, particularly
after taking into account that the survey strategy of PTF was far from
optimal, determined as it was by other science objectives, and probably at least a factor of two less effective than PanSTARRS. 

Also interesting are the statistics for other survey programs. 
For the period 1 May to 1 December 2014, Mt. Lemmon detected 21 NEAs with H>$27$ and the CSS detected 13.
DECam did most of its observations during other times of the year and had the highest number of detections of 
the smallest asteroids (H>28) over the entire year. 
The FoM for CSS in 2014 should be about 2.4 less than that for CSS-II shown in Table~\ref{tab:surveys}, due to its smaller FoV.  
The relatively high numbers of detections of NEAs with H$>$27 during the 1 May to 1 December 2014 period is likely due to the ability of CSS
to detect NEAs at less than the nominal 5$\sigma$ level used to calculate the FoM. 
\cite{christensen14} attributes this to the effectiveness of human scanning and validation of candidate NEAs.

Table~\ref{tab:surveys} also indicates the importance of accurate calibration of the efficiency of detecting
small NEOs if population size is to be estimated, particularly using peak detection.  
For the smallest fast-moving asteroids ($H=28.5$, $v_\text{ast}=20~{\rm km/s}$), there is typically a difference
of several hundred
between the effective detection volume employing streak detection and the point source detection volume,
i.e. same size asteroid but with $v_{\perp}=0$.
Calibration of the detection efficiency is therefore critical and can be accomplished using simulated NEO tracks 
injected into representative survey images (see\cite{allen+15} as an example).
Such a calibration can then be used to estimate the effective detection volume and a corresponding
population density of small NEOs.

\section*{Acknowledgements}

This work uses data obtained with the 1.2-m Samuel Oschin Telescope at Palomar Observatory 
as part of the Palomar Transient Factory project,
a scientific collaboration among the California Institute
of Technology, Columbia University, Las Cumbres Observatory, the Lawrence Berkeley National Laboratory, the
National Energy Research Scientific Computing Center,
the University of Oxford, and the Weizmann Institute of
Science; and the Intermediate Palomar Transient Factory
project,  a  scientific  collaboration  among  the  California
Institute  of  Technology,  Los  Alamos  National  Laboratory,  the  University  of  Wisconsin,  Milwaukee,  
the  Oskar Klein Center, the Weizmann Institute of Science, the
TANGO  Program  of  the  University  System  of  Taiwan,
and the Kavli Institute for the Physics and Mathematics
of the Universe.

The authors thank Robert Jedicke, and Quan-Zhi Ye for valuable comments on the paper, as well
as informative comments and suggestions by the referee.

A. Waszczak has been supported in part by the W.M. Keck Institute for Space Studies (KISS) at Caltech and this work was part of his thesis research. The work presented in this paper was in part inspired by a workshop on an Asteroid Retrieval Mission organized and funded by KISS in 2010--2012.  This work also was also supported by NASA JPL internal research and technology development funds.
Part of this research was carried out at the Jet Propulsion Laboratory,
California Institute of Technology, under a contract with the National
Aeronautics and Space Administration.

\appendix
\section{Figure of Merit for Detection of NEAs with Trailed Images}

In this appendix we derive a figure of merit (FoM) for detection of asteroids that are trailed in the
image.  We call this ``streak detection''. 
Elsewhere in the literature, this is sometimes called ``trail detection'' and the loss in detection 
sensitivity due to the trailed image is termed ``trailing loss''.
Here we take a somewhat different approach than some previous treatments (e.g.  
\cite{milani+96,rab91,jedicke+02,ver12}).  
We estimate the detection sensitivity loss in magnitudes and, in addition, use this
sensitivity loss to calculate a figure of merit by which to compare instruments with
different fields-of-view, limiting magnitude, exposure time, and readout time, taking into
account a distribution of asteroid velocity directions.

Assume a sky-background limited observation of a field. The
figure of merit (FoM) calculated here is a quantity proportional
to the average number of streaked asteroids detectable per unit time,
which for any given pointing direction is proportional to the detection
volume for asteroids of a given size and velocity perpendicular to the line of sight.
The FoM is defined to be:

\begin{equation}
FoM \equiv \frac{\Omega D^3}{\tau_{tot}}
\end{equation}

\noindent
and will depend on a number of factors including:
field of view, $\Omega$;
pixel size (arcsec), $\theta_\text{PSF}$;
limiting magnitude for detection of the telescope, $m_{lim}$;
the integration time of the observation, $\tau$, and the
total time for observations, $\tau_{tot}$, of a given field including number of revisits and dead time due to readout and slewing and
the velocity of the asteroid perpendicular to the line of sight, $v_\perp$.
We have chosen not to include the number of revisits in the definition of $FoM$ since
this can vary with time for a given survey.

\subsection{Point Source Detection FoM}

The distance, $D$, to which a point source can be detected is related to its magnitude, $m$, as:

\begin{equation*}
	m - m_0 = 5\  {\rm log}_{10} (D)
\end{equation*}
\noindent
where $m_0$ is a normalization.  Consequently
\begin{equation*}
	D = 10^{0.2(m - m_0)}
\end{equation*}
\noindent
The FoM  for point sources is then:

\begin{equation}
\text{FoM}_\text{pnt} = \frac{\Omega 10^{0.6(m_{lim} - m_0)}}{\tau_{tot}}
\label{eq:fom_pnt}
\end{equation}

\subsection{Streak Detection}

\subsubsection{Streak SNR}

The equations for a streak need to be modified 
because the signal-to-noise ($SNR$) saturates when the length of the streak becomes larger than $\theta_\text{PSF}$.  
While the number of pixels  containing a signal increases with time, the noise per pixel also increases, 
with the two effects canceling each other. 
Increases in exposure time therefore do not increase the effective SNR and the streak becomes more difficult
to detect because the surface brightness decreases with time.

The SNR has limiting cases:
\begin{align*}
SNR_\text{strk} &\sim \frac{F / (4 \pi D^2)}{\sqrt{B}} \times \sqrt{\tau_\text{PSF}}\ \ \ ({\rm for}\ \tau/\tau_\text{PSF} >> 1,{\rm \ i.e.\ streak})\\
SNR_\text{pnt}&\sim \frac{F / (4 \pi D^2)}{\sqrt{B}} \times {\sqrt{\tau}}\ \ \ ({\rm for}\ \tau/\tau_\text{PSF} << 1,{\rm \ i.e. \ point\ source})
\end{align*}  
\noindent
where  $F$ is the source strength, $B$ is the background flux per unit PSF, $\tau$ is the exposure time, $\tau_\text{PSF} = \theta_\text{PSF}/{\dot \theta}$
is the time required for the asteroid to traverse a length of one PSF,
and ${\dot \theta}$ is the angular velocity of the asteroid during the observation.

A useful approximation that has the proper limiting behaviors as well as appropriate intermediate behavior
(e.g. $SNR_\text{strk} <  SNR_\text{pnt}$ in all cases) is:

\begin{align}
SNR \sim\frac{F / (4 \pi D^2) } {\sqrt{ B}} \times \frac{ \sqrt{\tau}}{\sqrt{1+ \tau/\tau_\text{PSF}}} 
= \frac{F / (4 \pi D^2) } {\sqrt{ B}} \times \frac{ \sqrt{\tau}}{\sqrt{1+ \tau {\dot \theta}/\theta_\text{PSF}}}
\label{eq:snr_approx}
\end{align}

\subsubsection{Streak FoM}

The detection volume for a given type of source goes as $\Omega D^3/3$, where $D$ is the detection limit of the telescope for that source.
The detection limit is defined for some limiting SNR, say $SNR=5$.  
We assume that the density of asteroids and their velocity distribution is independent of distance (a reasonable assumption
as most are detected fairly close to Earth).
To achieve an SNR of 5 for a streaked object, the object must be closer than a point source of the same magnitude.
Specifically, because $SNR^2 \propto D^{-4}$, the decrease in distance to achieve the same SNR as a point source implies
(using Eq. \ref{eq:snr_approx}):

\begin{equation}
	D^4_\text{strk}  \sim D^4_\text{pnt} \times  \frac{1}{1+\tau {\dot \theta}/\theta_\text{PSF}}
\label{eq:Dstrk}
\end{equation}

This yields an expression for the decrease in limiting magnitude due to a streak of:

\begin{equation}
	\Delta m_\text{lim} = 5 \log_{10} \left( D_\text{pnt}/D_\text{strk} \right) \sim 1.25 \log_{10} \left( \frac{1}{1+\tau {\dot \theta}/\theta_\text{PSF}} \right)
\label{eq:maglength}
\end{equation}

\noindent
Note that the reduction in limiting magnitude in Eq. \ref{eq:maglength} is not the same as the trailing loss often discussed in the literature. 
The magnitude change  in Eq. \ref{eq:maglength} is the decrease in limiting magnitude assuming that a candidate streak has been identified 
and that the signals in the pixels along
the streak have been summed together, i.e. what we call streak detection. 
This is the same as the quantity ``SNR trailing loss'' referred to in the context of LSST by \cite{jones+16}.
As will be shown in Section \ref{sec:peak}, the magnitude change in Eq. \ref{eq:maglength} is half of what is usually called trailing loss.
Indeed, the gain in detection rate using streak detection is a direct result of the fact that the magnitude losses are only half that incurred in 
conventional ``peak detection''.

\begin{figure}
\caption{\small Computed limiting magnitude vs streak length.}
\begin{center}
\vspace{.2in}
\centerline {
\includegraphics[width=4in]{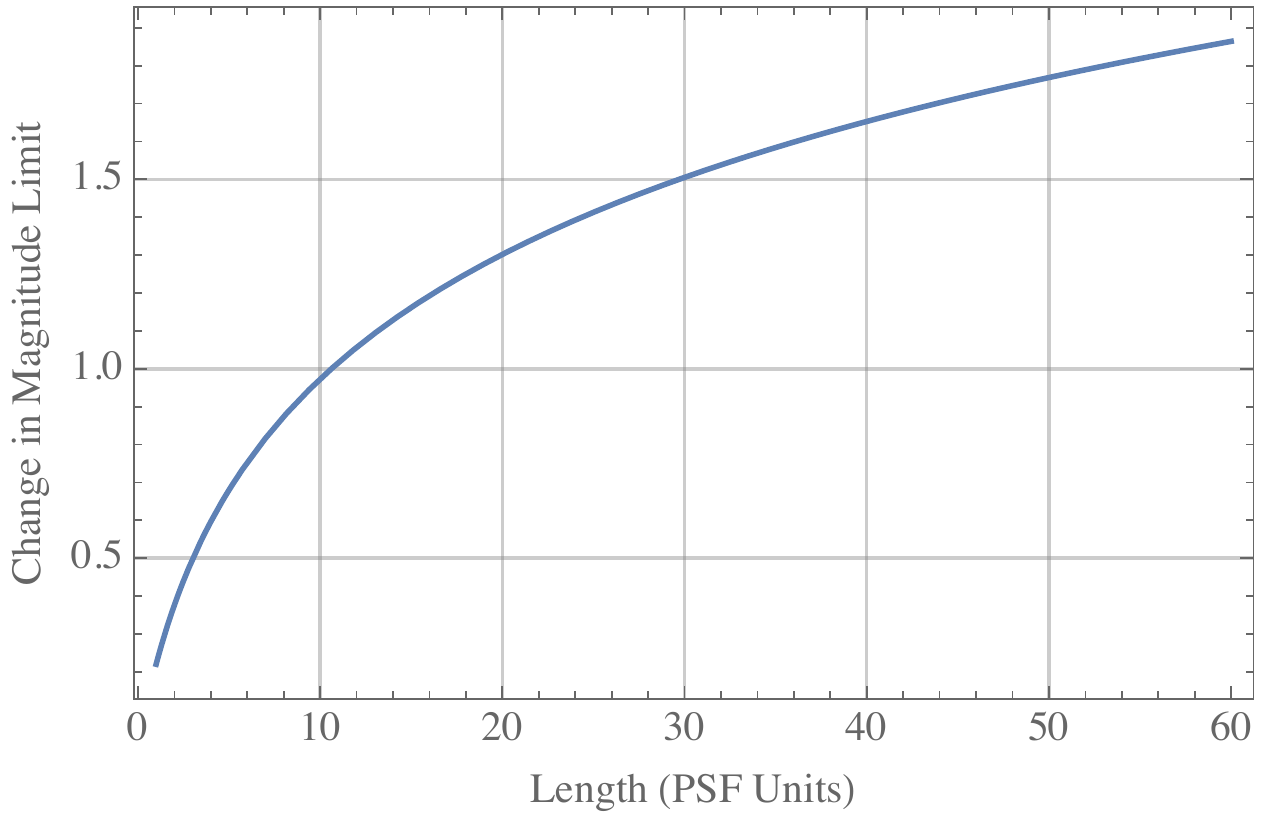}
}
\vspace{.2in}
\label{fig:maglimvslength}
\end{center}
\end{figure}

Fig. \ref{fig:maglimvslength} shows the estimated magnitude limit versus streak length in PSF units as computed from
Eq. \ref{eq:maglength}.

Returning to the derivation of a figure of merit and dropping the approximate symbols, Eq. \ref{eq:Dstrk} can be expressed as:

\begin{align}
	D^4_\text{strk} &= D^4_\text{pnt} \times  \frac{1}{1+\tau {\dot \theta}/ \theta_\text{PSF}}\\
	\label{eq:FOMfull} &= D^4_\text{pnt} \times  \frac{1}{1+\tau (v_\perp/D_\text{strk}) / \theta_\text{PSF}}\\
	&\approx D^4_\text{pnt} \times  \frac {D_\text{strk} \theta_\text{PSF}}{\tau v_\perp}  {\ \ \ \ \ \ \ \rm (for\  \tau {\dot \theta} \gg \theta_\text{PSF})}
\end{align}
\noindent
where $v_\perp$ is the asteroid velocity perpendicular to the line of sight, and therefore

\begin{align}
	D_\text{strk}^3 &\approx  D^3_\text{pnt} \times  \frac  { D_\text{pnt} \theta_\text{PSF}}{\tau v_\perp}  {\ \ \ \ \ \ \ \rm (for\  \tau {\dot \theta} \gg \theta_\text{PSF})}
	\label{eq:FoMapprox}
\end{align}

Using Eq. \ref{eq:FoMapprox} then yields for the FoM:

\begin{align}
	\text{FoM}_\text{strk} &= \frac{\Omega D_\text{strk}^3} {\tau_{tot}} 
	\approx \frac{\Omega \ D_\text{pnt}^3}{\tau_{tot}} \left( \frac{D_\text{pnt} \theta_\text{PSF}}{\tau v_\perp} \right) 
	= \text{FoM}_\text{pnt} \left( \frac{ \theta_\text{PSF}}{ \theta_\text{strk}} \right)  {\ \ \ \ \ \ \ \rm (for\  \tau {\dot \theta} \gg \theta_\text{PSF})}
	\label{eq:FOM1}
\end{align}

\noindent
where $\tau_{tot}$ is the total time per field including integration time, readout time, and slew time, and $\theta_\text{strk}$ is the angular extent of the streak
\emph{at the point source limiting distance for the asteroid}.
Note the difference between $v_\perp$ and $v_\text{ast}$, as used in Section \ref{sec:scaling}.  
If we integrate Eq. \ref{eq:FOM1} over an isotropic distribution of asteroid velocity directions,
then we obtain an additional $\pi/2 \sim 1.57$ in the volume factor.


The approximation $\tau {\dot \theta} \gg \theta_\text{PSF}$ in Eq. \ref{eq:FoMapprox}  is not strictly necessary. 
The formula for $D_\text{strk}$
(Eq. \ref{eq:FOMfull}) 
can be solved exactly, although the expression is rather cumbersome.
Mathematica was used to derive the exact solution and it is used in the calculations of FoM.
The exact expression requires choosing a fiducial $v_\perp$.  
An advantage to the exact calculation is that there is a smooth transition between point source behavior
and streaked source behavior.
The averaging over asteroid velocity directions can be done for the exact calculation, but involves
a numerical evaluation. 
Results of these calculations are given in Table~\ref{tab:surveys}.

\subsection{Peak Detection}
\label{sec:peak}

To estimate the detection volumes for a survey that does not specifically implement streak detection,
we estimate the probability that the measured flux exceeds the limiting magnitude at some point along the streak.
We call this peak detection.
For peak detection within a streak, the surface brightness of the streak decreases like $\tau/\tau_\text{PSF}$ and therefore
in analogy to Eq. \ref{eq:snr_approx}:

\begin{align*}
SNR \sim\frac{F / (4 \pi D^2) } {\sqrt{ B}} \times \frac{ \sqrt{\tau}}{{1+ \tau {\dot \theta}/\theta_\text{PSF}}}
\label{eq:snr_peak_approx}
\end{align*}

Consequently, the detection distance for peak and point detection are related by:

\begin{equation}
	D^2_\text{peak}  = D^2_\text{pnt} \times  \frac{1}{1+\tau {\dot \theta}/\theta_\text{PSF}}
\label{eq:Dpeak}
\end{equation}

This yields an expression for the decrease in limiting magnitude due to streak length of:

\begin{equation}
	\Delta m_{lim} = 5 \log_{10} \left( D_\text{pnt}/D_\text{peak} \right) = 2.5 \log_{10} \left( \frac{1}{1+\tau {\dot \theta}/\theta_\text{PSF}} \right)
	\label{eq:magpeak}
\end{equation}

\noindent
i.e. the decrease in the magnitude limit for simple peak detection is double that of full streak detection.
This change in magnitude (Eq. \ref{eq:magpeak}) is often called the ``trailing loss'' in the literature. 
As examples, equation \ref{eq:magpeak} accurately describes the measured loss in sensitivity shown in Figure~2 of \cite{jedicke+herron97} for the case of
trailed images in Spacewatch, and also agrees to within about 20\% with the estimate of trailing losses for LSST.
See the upper curve in Figure~5 of \cite{jones+16}). 
Equations \ref{eq:maglength} and \ref{eq:magpeak} provide a natural explanation of the factor of two difference between the ``SNR trailing losses'' and 
``detection trailing losses'', shown in that figure.

Eq. \ref{eq:Dpeak} can be expanded to yield:
\begin{align*}
	D^2_\text{peak}  &= D^2_\text{pnt} \times  \frac{1}{1+\tau {\dot \theta}/ \theta_\text{PSF}}\\
	&= D^2_\text{pnt} \times  \frac{1}{1+\tau (v_\perp) /( \theta_\text{PSF} D_\text{peak})}
	\label{eq:FoMpeak}
\end{align*}

This can be solved exactly to yield: 
\begin{align}
	D_\text{peak} = \frac{1}{2} \left( \sqrt{4 D_\text{pnt}^2 + \left( \frac{\tau v_\perp}{\theta_\text{PSF}}\right)^2} - \frac{\tau v_\perp}{\theta_\text{PSF}}\right)
\end{align}

which can then be inserted into the FoM:

\begin{align}
	\text{FoM}_\text{peak} \equiv \frac{\Omega D_\text{peak}^3} {\tau_{tot}} \left(10^{ \Delta m_\text{multi}}\right)
	\label{eq:FOM1_peak}
\end{align}

\noindent
where we account for the small
improvement, $\Delta m_\text{multi}$, in effective limiting magnitude due to the fact that there are multiple possibilities for detection along the streak, thus
increasing the probability for exceeding the detection threshold.
The averaging over angles can also be done for the peak case, and requires a numerical integration over the effective
volume for detection as a function of $v_\perp$. 
The results of these numerical computations of the peak detection FoM are provided in Table~\ref{tab:surveys} as numbers
in parentheses.

\subsection{Discovery vs Detection}

The figures of merit discussed in this appendix relate to detection rather than discovery of new objects.
That is, a figure of merit such as $\Omega D^3/\tau_{tot}$ is proportional to the rate of detection of asteroids, with
no distinction as to whether the asteroid has been previously detected or not.
An estimate of the discovery rate requires an estimate of the detection rate of new asteroids and involves several
factors: the number density of asteroids of a given type as a function of direction, the rate of change of the population
of asteroids in the detection volume of a given survey, and the time for a given survey to cover the observable sky.

Two time scales are important: the time scale for covering the observable sky, $\tau_\text{cover}$, and the typical
residence time of an asteroid in the detection volume, $\tau_\text{res}$.  
If $\tau_\text{res} > \tau_\text{cover}$, then a given asteroid may be detected more than once by a given survey
as it surveys the sky multiple times.
In this case, the rate of detection of new asteroids will be governed by the rate of change of the population within the detection volume.
This is the typical case for large asteroids for which the detection volume is large.
In the opposite case, $\tau_\text{res} < \tau_\text{cover}$, the population within the detection volume changes more 
rapidly than the detection volume can be surveyed, and the rate of detection of new asteroids is governed by the area
coverage rate of the survey. 
This is the typical case for very small asteroids for which the detection volume is small.
Because $\tau_\text{res}$ depends on the size of the detection volume, i.e. the depth of the survey, while $\tau_\text{cover}$
depends on the rate of areal coverage, the rate of new asteroid detection will be a strong function of asteroid size 
and will also depend strongly on whether a survey is shallow with a wide field of view, or deep with a narrow field of view.
A detailed discussion of these factors is beyond the scope of this appendix.

\end{document}